\documentclass{article}

\pdfoutput=1
\usepackage{amssymb}
\usepackage{amsfonts}
\usepackage{mathrsfs}
\usepackage{bbm}
\usepackage{amsthm}
\usepackage{fontenc}
\usepackage{latexsym}
\usepackage{amsmath}
\usepackage{amsxtra}
\usepackage{amstext}
\usepackage{txfonts}
\usepackage{stmaryrd}
\usepackage{accents}
\usepackage{upgreek}
\usepackage{graphicx}
\usepackage{overpic}

\usepackage{textcomp}

\usepackage{verbatim}

\theoremstyle{plain}
\newtheorem{theorem}{Theorem}[section]

\newtheorem{lemma}[theorem]{Lemma}
\newtheorem{remark}{Remark}[section]

\newcommand{\secref}[1]{\S\ref{#1}}
\newcommand{\lemref}[1]{Lemma~\ref{#1}}

\renewcommand{\sectionmark}[5]{}

\def\c{\mathbb{C}}
\def\as{\text{ as }}

\newcommand{\re}{\operatorname{Re}}

\begin{document}
\begin{center}
\textbf{\Large Uniqueness of Kerr-Newman solution} \\

\vspace{1pt}
A.K.M. Masood-ul-Alam \\

Mathematical Sciences Center and Department of Mathematical
Sciences, Tsinghua University \\ Haidian District, Beijing 100084,
PRC \\ abulm@math.tsinghua.edu.cn

\end{center}

\vspace{1pt}

\textbf{Abstract. \ }  We show that non-degenerate multiple
black hole solution of Einstein-Maxwell equations in an
asymptotically flat axisymmetric spacetime cannot be in stationary
equilibrium. This extends the uniqueness of Kerr-Newman solution
first proved by Bunting and Mazur in a much wider desirable class.
Spin-spin interaction cannot hold the black hole aparts even with
electromagnetic forces.

\vspace{1pt}


\section{Introduction}

We generalize the method used in \cite{MA1},\cite{MA2} for the
uniqueness problem of Kerr-Newman solution for
$M^{2}>a^{2}+\mathfrak{e}^{2}+\mathfrak{m}^{2}.$ Here
$\mathfrak{m}$ is the magnetic charge. For a single black hole
case the results are due to Bunting \cite{Bunt} and Mazur
\cite{Maz} (also Carter \cite{Cbm}) using different techniques. Several extensions of these results have been obtained by
 Wells \cite{Wel}. Regarding the possibility of multiple black holes including those in a vacuum spacetime several
 results are obtained
by Weinstein \cite{9,10} (see also the review article by Beig and
Chrusciel \cite{11}), Neugebauer and Meinel \cite{12}, Chrusciel
and Costa \cite{13}, Wong and Yu
\cite{14}. Wong and Yu does not need the axisymmetric assumption but assumed the solution to be close to Kerr-Newman solution in
some sense. We consider only non-degenerate black hole boundary.
Our technique involves tailoring suitably the spinorial proofs of
the positive mass theorem of Schoen and Yau
\cite{SY} due to Witten
\cite{W} and Bartnik
\cite{B}.

Initially we suppose the spacetime metric is stationary and
axisymmetric EM black hole solution having the form
\begin{equation} \label{1}
\overset{4}{g}=\overset{4}{g}_{ab}dx^{a}dx^{b}=-Vdt^{2}+2Wdtd\phi
+Xd\phi ^{2}+\overline{g}
\end{equation}
 Carter (Part II, \cite{C1}) showed that $\overline{g}=\Omega (d\rho
^{2}+dz^{2})$ where $\rho $ and $z$ are conjugate harmonic
functions. $V,$ $W,$ $X$ and $\Omega$ are functions of $\rho $ and
$z.$ Following Carter we take
\begin{equation} \label{2}
W^{2}+VX=\rho ^{2}
\end{equation}

\section{Ricci Curvature}

We denote the 2-metric by
$\overline{g}=\overset{4}{g}_{11}(dx^{1})^{2}+
\overset{4}{g}_{22}(dx^{2})^{2}=\overset{4}{g}_{AB}dx^{A}dx^{B}.$
We denote the induced 3-metric on a $t=\text{constant}$ hypersurface by
\begin{equation} \label{3}
\widehat{g}=\overline{g}+Xd\phi ^{2}
\end{equation}
Then $\det ([\overset{4}{g}_{ab}])=-\left( VX+W^{2}\right)
\overline{g}_{11}\overline{g}_{22}.$ Using Eq.~\eqref{2} we get
\scriptsize
 $$[\overset{4}{ g^{ab}}]=
\begin{bmatrix}
-X\rho^{-2} & 0 & 0 & W\rho^{-2} \\ 0 & \overline{g}_{11}^{-1} & 0
& 0 \\ 0 & 0 & \overline{g}_{22}^{-1} & 0 \\ W\rho^{-2} & 0 & 0 &
V\rho^{-2}
\end{bmatrix}
$$
\normalsize
 Henceforth $\overline{g}^{AB}$ is obtained from
$\overline{g}_{AB}$ by raising the indices with the 2-metric
$\overline{g}.$ The $t$=constant surface has zero mean curvature.
$\widehat{g}$ is a Riemannian metric. We shall denote the
Laplacian and the covariant derivative of the two-metric
$\overline{g}$ by $\overline{\Delta }$ and $\overline{\nabla
\strut }.$ In general we shall use the
metrics as subscripts in order to indicate w.r.t. which metric a
norm or an operator is computed. Since we shall not use the usual
formulations of a stationary axisymmetric vacuum spacetime it is
better to give the expressions for the components of the Ricci
curvature $\overset{4}{g}$ for easy reference before equating them
to zero using the vacuum Einstein equations. Ricci curvature of
the four metric is
\scriptsize
\begin{eqnarray}
\overset{4}{R}_{tt} &=&\dfrac{1}{2}\overline{\Delta }V+\dfrac{V}{4\rho ^{2}}
\left\langle \overline{\nabla \strut }V,\overline{\nabla \strut }
X\right\rangle +\dfrac{V}{2\rho ^{2}}|\overline{\nabla \strut
}W|^{2}-\dfrac{ W}{2\rho ^{2}}\left\langle \overline{\nabla \strut
}V,\overline{\nabla
\strut }W\right\rangle -\dfrac{X}{4\rho ^{2}}|\overline{\nabla \strut
}V|^{2} \label{rtt}
\\
\overset{4}{R}_{t\phi } &=&-\dfrac{1}{2}\overline{\Delta }W+\dfrac{V}{4\rho
^{2}}\left\langle \overline{\nabla \strut }W,\overline{\nabla
\strut } X\right\rangle -\dfrac{W}{2\rho ^{2}}\left\langle
\overline{\nabla \strut }V,
\overline{\nabla \strut }X\right\rangle +\dfrac{X}{4\rho ^{2}}\left\langle
\overline{\nabla \strut }V,\overline{\nabla \strut }W\right\rangle \label{rtphi} \\
\overset{4}{R}_{\phi \phi } &=&-\dfrac{1}{2}\overline{\Delta }X-\dfrac{X}{
4\rho ^{2}}\left\langle \overline{\nabla \strut
}V,\overline{\nabla \strut } X\right\rangle +\dfrac{W}{2\rho
^{2}}\left\langle \overline{\nabla \strut }W,
\overline{\nabla \strut }X\right\rangle -\dfrac{X}{2\rho ^{2}}|\overline{
\nabla \strut }W|^{2}+\dfrac{V}{4\rho ^{2}}|\overline{\nabla \strut
}X|^{2} \label{rphiphi}
\\
R_{At}^{(4)} &=&0=R_{A\phi }^{(4)} \label{rAtphi} \\
\overset{4}{R}_{BD} &=&\dfrac{1}{2}\overline{R}\overline{g}_{BD}-\dfrac{1}{
\rho }\overline{\nabla \strut }_{D}\overline{\nabla \strut }_{B}\rho +\dfrac{
1}{2\rho ^{2}}\overline{\nabla \strut }_{D}W\overline{\nabla
\strut }_{B}W+
\dfrac{1}{4\rho ^{2}}\overline{\nabla \strut }_{D}V\overline{\nabla \strut }
_{B}X+\dfrac{1}{4\rho ^{2}}\overline{\nabla \strut
}_{B}V\overline{\nabla
\strut }_{D}X \label{rAB}
\end{eqnarray}
\normalsize
\section{Einstein equation}
\begin{equation}\label{einstein}
  \overset{4}{R}_{ab}=8\pi\left(T_{ab}-(1/2)T\overset{4}{g}_{ab}\right)
\end{equation}
with energy-momentum tensor
\begin{equation}\label{enm}
  T_{ab}=\dfrac{1}{4\pi}\left(F_{ac}F_{bd}\overset{4}{g}^{cd}-(1/4)\overset{4}{g}_{ab}F_{ij}F_{kl}\overset{4}{g}^{ik}\overset{4}{g}^{jl}\right)
\end{equation}
F is the electromagnetic field tensor obtained from the
electromagnetic potential one form $\mathbf{A}$
\begin{equation}\label{F}
F_{ab}=\dfrac{\partial
\mathbf{A}_{b}}{\partial
x^{a}}-\dfrac{\partial
\mathbf{A}_{a}}{\partial
x^{b}}
\end{equation}
Since $T=\overset{4}{g}^{ab}T_{ab}=0$ Einstein equation becomes
\begin{equation}\label{einstein}
  \overset{4}{R}_{ab}=8\pi T_{ab}
\end{equation}

Following Carter (Eq.~7.43 Part II \cite{C1}) we choose the
electromagnetic potential to be of the form
\begin{equation}\label{a}
\mathbf{A}=\xi dt+\psi d\phi
\end{equation}
where $\psi$ is a function of $x^{1}$ and $x^{2}.$ (Our sign in
Eq.~\eqref{F} is opposite to that of Eq.6.48 in Carter). Thus

One finds
\begin{eqnarray}
  T_{tt}=&\dfrac{1}{8\pi}\left(|\overline{\nabla\strut}
\xi|^{2}+\left|W\overline{\nabla\strut}
\xi+V\overline{\nabla\strut}
\psi\right|^{2}\rho^{-2}\right)\\
  T_{t\phi}=& \dfrac{1}{8\pi}\left(2VX\left\langle\overline{\nabla
\strut}\xi,\overline{\nabla
\strut}\psi\right\rangle\rho^{-2}+WX|\nabla\strut
\xi|^{2}\rho^{-2}
-WV|\overline{\nabla\strut}
\psi|^{2}\rho^{-2}\right)\\
  T_{\phi\phi} =& \dfrac{1}{8\pi}\left(|\overline{\nabla\strut}
\psi|^{2}+\left|X\overline{\nabla\strut}
\xi+W\overline{\nabla\strut}
\psi\right|^{2}\rho^{-2}\right) \\
  T_{AB} =& \dfrac{1}{4\pi\rho^{2}}\left(-X\dfrac{\partial
\xi}{\partial
x^{A}}\dfrac{\partial
\xi}{\partial
x^{B}}+W\dfrac{\partial
\xi}{\partial
x^{A}}\dfrac{\partial
\psi}{\partial
x^{B}}+W\dfrac{\partial
\psi}{\partial
x^{A}}\dfrac{\partial
\xi}{\partial
x^{B}} + V\dfrac{\partial
\psi}{\partial
x^{A}}\dfrac{\partial
\psi}{\partial
x^{B}}\right.\\ &\left.+\dfrac{X}{2}|\overline{\nabla\strut}
\xi|^{2}\overline{g}_{AB}- W\left\langle\overline{\nabla
\strut}\xi,\overline{\nabla
\strut}\psi\right\rangle\overline{g}_{AB}
-\dfrac{V}{2}|\overline{\nabla\strut}
\psi|^{2}\overline{g}_{AB}\right)\\
T_{At} =&0=T_{A\phi}
\end{eqnarray}

Einstein equation Eq.~\eqref{einstein} becomes
\begin{eqnarray}
\overset{4}{R}_{tt}=&|\overline{\nabla\strut}
\xi|^{2}+\left|W\overline{\nabla\strut}
\xi+V\overline{\nabla\strut}
\psi\right|^{2}\rho^{-2}
\\ \label{riccien1}
\overset{4}{R}_{t\phi }=&2VX\left\langle\overline{\nabla
\strut}\xi,\overline{\nabla
\strut}\psi\right\rangle\rho^{-2}+WX|\nabla\strut
\xi|^{2}\rho^{-2}
-WV|\overline{\nabla\strut}
\psi|^{2}\rho^{-2}
\\ \label{riccien2}
\overset{4}{R}_{\phi \phi }=&|\overline{\nabla\strut}
\psi|^{2}+\left|X\overline{\nabla\strut}
\xi+W\overline{\nabla\strut}
\psi\right|^{2}\rho^{-2}
\\ \label{riccien3}
\overset{4}{R}_{BD}=&\dfrac{2}{\rho^{2}}\left(-X\dfrac{\partial
\xi}{\partial
x^{B}}\dfrac{\partial
\xi}{\partial
x^{D}}+W\dfrac{\partial
\xi}{\partial
x^{B}}\dfrac{\partial
\psi}{\partial
x^{D}}+W\dfrac{\partial
\psi}{\partial
x^{B}}\dfrac{\partial
\xi}{\partial
x^{D}} + V\dfrac{\partial
\psi}{\partial
x^{B}}\dfrac{\partial
\psi}{\partial
x^{D}}\right.\\ &\left.+\dfrac{X}{2}|\overline{\nabla\strut}
\xi|^{2}\overline{g}_{BD}- W\left\langle\overline{\nabla
\strut}\xi,\overline{\nabla
\strut}\psi\right\rangle\overline{g}_{BD}
-\dfrac{V}{2}|\overline{\nabla\strut}
\psi|^{2}\overline{g}_{BD}\right) \label{riccien4}
\end{eqnarray}

First three of the following five equations we get from
Eqs.~(\ref{rtt}-\ref{rphiphi}) and
Eqs.~(\ref{riccien1}-\ref{riccien3}) respectively. Last two
equations are the non-trivial equations in the nontrivial set of
Maxwell equations namely
$F_{ab}\overset{4}{;}_{c}\overset{4}{g}^{bc}=0$ for $a=0$ and
$a=3.$
\scriptsize
\begin{eqnarray}
\overline{\Delta }V &=&-\dfrac{V}{2\rho ^{2}}\left\langle \overline{\nabla
\strut }X,\overline{\nabla \strut }V\right\rangle -\dfrac{V|\overline{\nabla
\strut }W|^{2}}{\rho ^{2}}+\dfrac{W}{\rho ^{2}}\left\langle \overline{\nabla
\strut }W,\overline{\nabla \strut }V\right\rangle +\dfrac{X|\overline{\nabla
\strut }V|^{2}}{2\rho ^{2}}+2|\overline{\nabla\strut}
\xi|^{2}+2\left|W\overline{\nabla\strut}
\xi+V\overline{\nabla\strut}
\psi\right|^{2}\rho^{-2} \label{LV} \\
\overline{\Delta }W&=&\dfrac{V}{2\rho ^{2}}\left\langle \overline{\nabla
\strut }X,\overline{\nabla \strut }W\right\rangle -\dfrac{W}{\rho ^{2}}
\left\langle \overline{\nabla \strut }V,\overline{\nabla \strut }
X\right\rangle +\dfrac{X}{2\rho ^{2}}\left\langle \overline{\nabla
\strut }V,
\overline{\nabla \strut }W\right\rangle-\dfrac{4VX}{\rho^{2}}\left\langle\overline{\nabla
\strut}\xi,\overline{\nabla
\strut}\psi\right\rangle
+\dfrac{2W}{\rho^{2}}\left(V|\overline{\nabla\strut}
\psi|^{2}-X|\nabla\strut
\xi|^{2}\right) \label{LW} \\
\overline{\Delta }X &=&-\dfrac{X}{2\rho ^{2}}\left\langle \overline{\nabla
\strut }V,\overline{\nabla \strut }X\right\rangle +\dfrac{W}{\rho ^{2}}
\left\langle \overline{\nabla \strut }W,\overline{\nabla \strut }
X\right\rangle -\dfrac{X|\overline{\nabla \strut }W|^{2}}{\rho
^{2}}+\dfrac{ V|\overline{\nabla \strut }X|^{2}}{2\rho ^{2}}
 \label{LX}-2|\overline{\nabla\strut}
\psi|^{2}-2\left|X\overline{\nabla\strut}
\xi+W\overline{\nabla\strut}
\psi\right|^{2}\rho^{-2} \\
\overline{\Delta}\xi &=&\dfrac{X}{2\rho^{2}}\left\langle \overline{\nabla\strut}V,\overline{\nabla\strut}\xi\right\rangle
- \dfrac{V}{2\rho^{2}}\left\langle
\overline{\nabla\strut}X,\overline{\nabla\strut}\xi\right\rangle+
\dfrac{V}{\rho^{2}}\left\langle
\overline{\nabla\strut}W,\overline{\nabla\strut}\psi\right\rangle-
\dfrac{W}{\rho^{2}}\left\langle
\overline{\nabla\strut}V,\overline{\nabla\strut}\psi\right\rangle
\\ \label{Lxi}
  \overline{\Delta}\psi
&=&- \dfrac{X}{2\rho^{2}}\left\langle
\overline{\nabla\strut}V,\overline{\nabla\strut}\psi\right\rangle
  +\dfrac{V}{2\rho^{2}}\left\langle \overline{\nabla\strut}X,\overline{\nabla\strut}\psi\right\rangle
-\dfrac{X}{\rho^{2}}\left\langle
\overline{\nabla\strut}W,\overline{\nabla\strut}\xi\right\rangle+
\dfrac{W}{\rho^{2}}\left\langle
\overline{\nabla\strut}X,\overline{\nabla\strut}\xi\right\rangle \label{Lpsi}
\end{eqnarray}
\normalsize
Last two equations are the same as in Bunting's thesis (replacing
his functions $E,F,A,B,C$ by $\xi,\psi,-V,X,W$) and are equivalent
to the set given by Carter (p74, Part II \cite{C1}):
\begin{eqnarray*}
\overline{\nabla \strut}\left(\left(X\overline{\nabla \strut}\xi-W\overline{\nabla
  \strut}\psi\right)\rho^{-1}\right)=0 \\
  \overline{\nabla \strut}\left(\rho X^{-1}\overline{\nabla \strut}\psi
  +W(\rho X)^{-1}\left(X\overline{\nabla \strut}\xi-W\overline{\nabla
  \strut}\psi\right)\right)=0
\end{eqnarray*}

From Eqs.~(\ref{LV}-\ref{LX}) one can show that $\rho
=\sqrt{VX+W^{2}}$ is a harmonic function i.e. $\overline{\Delta
}\rho =0.$ This we are assuming from the start.

\section{Remaining equations}

Eqs.~(\ref{rAB},\ref{riccien4}) give
\scriptsize
\begin{multline}\label{rBD1a}
\dfrac{1}{2}\overline{R}\overline{g}_{BD}-\dfrac{1}{
\rho }\overline{\nabla \strut }_{D}\overline{\nabla \strut }_{B}\rho +\dfrac{
1}{2\rho ^{2}}\dfrac{\partial W}{\partial x^{D}}\dfrac{\partial
W}{\partial x^{B}}+
\dfrac{1}{4\rho ^{2}}\dfrac{\partial V}{\partial x^{D}}\dfrac{\partial
X}{\partial x^{B}}+\dfrac{1}{4\rho ^{2}}\dfrac{\partial
V}{\partial x^{B}}\dfrac{\partial X}{\partial
x^{D}}\\=\dfrac{2}{\rho^{2}}\left(-X\dfrac{\partial
\xi}{\partial
x^{B}}\dfrac{\partial
\xi}{\partial
x^{D}}+W\dfrac{\partial
\xi}{\partial
x^{B}}\dfrac{\partial
\psi}{\partial
x^{D}}+W\dfrac{\partial
\psi}{\partial
x^{B}}\dfrac{\partial
\xi}{\partial
x^{D}} + V\dfrac{\partial
\psi}{\partial
x^{B}}\dfrac{\partial
\psi}{\partial
x^{D}}+\dfrac{X}{2}|\overline{\nabla\strut}
\xi|^{2}\overline{g}_{BD}- W\left\langle\overline{\nabla
\strut}\xi,\overline{\nabla
\strut}\psi\right\rangle\overline{g}_{BD}
-\dfrac{V}{2}|\overline{\nabla\strut}
\psi|^{2}\overline{g}_{BD}\right)
\end{multline}
\normalsize
Contracting we get
\begin{equation}\label{R2a}
  \overline{R}=-\dfrac{
1}{2\rho ^{2}}|\overline{\nabla \strut }W|^{2}-
\dfrac{1}{2\rho ^{2}}\left\langle\overline{\nabla \strut }V,\overline{\nabla \strut }
X\right\rangle
\end{equation}
\normalsize
Differentiating Eq.~\eqref{2} we get $2\rho\overline{\nabla \strut
}
\rho =V\overline{\nabla \strut }X+X\overline{\nabla
\strut }V+2W\overline{\nabla \strut }W $ so that
\begin{align}
2\rho \left\langle \overline{\nabla \strut }\rho ,\overline{\nabla
\strut } V\right\rangle &=V\left\langle \overline{\nabla
\strut }X,\overline{\nabla \strut }V\right\rangle +2W\left\langle \overline{
\nabla \strut }W,\overline{\nabla \strut }V\right\rangle +X|\overline{\nabla
\strut }V|^{2} \label{8} \\
2\rho \left\langle \overline{\nabla \strut }\rho ,\overline{\nabla
\strut } X\right\rangle &= X\left\langle \overline{\nabla
\strut }V,\overline{\nabla \strut }X\right\rangle +2W\left\langle \overline{
\nabla \strut }W,\overline{\nabla \strut }X\right\rangle +V|\overline{\nabla
\strut }X|^{2} \label{9} \\
2\rho \left\langle \overline{\nabla \strut }\rho ,\overline{\nabla
\strut } W\right\rangle &= 2W\overline{|\nabla \strut }
W|^{2}+V\left\langle \overline{\nabla \strut }X,\overline{\nabla
\strut } W\right\rangle +X\left\langle \overline{\nabla \strut
}V,\overline{\nabla
\strut }W\right\rangle  \label{10}
\end{align}

Using the above relations and Eq.~\eqref{R2a} we write
Eqs.~(\ref{LV},\ref{LX}) as
\begin{align}
\overline{\Delta }V &=2\overline{R}V+\left\langle \overline{\nabla \strut }
\ln \rho ,\overline{\nabla \strut }V\right\rangle+2|\overline{\nabla\strut}
\xi|^{2}+2\left|W\overline{\nabla\strut}
\xi+V\overline{\nabla\strut}
\psi\right|^{2}\rho^{-2}  \label{lv} \\
\overline{\Delta }W &=2\overline{R}W+\left\langle \overline{\nabla \strut }
\ln \rho ,\overline{\nabla \strut }W\right\rangle -\dfrac{4VX}{\rho^{2}}\left\langle\overline{\nabla
\strut}\xi,\overline{\nabla
\strut}\psi\right\rangle
+\dfrac{2W}{\rho^{2}}\left(V|\overline{\nabla\strut}
\psi|^{2}-X|\nabla\strut
\xi|^{2}\right)\label{lw} \\
\overline{\Delta }X &=2\overline{R}X+\left\langle \overline{\nabla \strut }
\ln \rho ,\overline{\nabla \strut }X\right\rangle -2|\overline{\nabla\strut}
\psi|^{2}-2\left|X\overline{\nabla\strut}
\xi+W\overline{\nabla\strut}
\psi\right|^{2}\rho^{-2} \label{lx}
\end{align}
\normalsize
Since it is well-known that the equations for $\Omega$ can be
solved using its asymptotic value once we know the other functions
we do not include the complicated equations for it.

The $t=\text{constant}$ hypersurface has the topology
$\Sigma^{+}\cup\partial \Sigma^{+}$ where $\Sigma^{+}$ is an open
3-manifold and the boundary $\partial \Sigma^{+}$ is a finite
number of disconnected 2-spheres. $X>0$ in $\Sigma^{+}$ except on
the axis. $(\partial
\Sigma^{+},\overline{g})$ is a smooth totally geodesic submanifold
of the 3-dimensional Riemannian manifold with boundary
$(\Sigma^{+}\cup\partial \Sigma^{+},\widehat{g}).$ the 3-metric
$\widehat{g}$ has nonnegative scalar curvature, which is easy to
see from Eq.~\eqref{einstein}, weak energy condition and doubly
contracted Gauss equation for the maximal $t=\text{constant}$
hypersurface.

Let $\varrho ^{2}= r^{2}+a^{2}\cos ^{2}\theta$ and
\begin{equation}\label{varmass}
  M^{\prime}=M-\dfrac{\mathfrak{e}^{2}+\mathfrak{m}^{2}}{2r}
\end{equation}
 Kerr-Newman solution  has the spacetime metric,
\begin{equation} \label{kn}
\begin{split}
ds^{2}=-\left(1-\dfrac{2M^{\prime}r}{\varrho
^{2}}\right)dt^{2}-\dfrac{4M^{\prime}ra\sin ^{2}\theta }{
\varrho ^{2}}d\phi dt+&\left( (r^{2}+a^{2})\sin ^{2}\theta +\dfrac{
2M^{\prime}ra^{2}\sin ^{4}\theta }{\varrho ^{2}}\right) d\phi ^{2}
\\ &+\varrho ^{2}\left(
\dfrac{dr^{2}}{r^{2}-2M^{\prime}r+a^{2}}+d\theta ^{2}\right)
\end{split}
\end{equation}
Soon after Kerr's discovery \cite{Kerr}, Kerr-Newman solution was
found by Newman and {\it et al.}
\cite{Newman}. The form given above is obtained from Carter (Eq.~5.54, Part I, \cite{C1}) by collecting the
terms containing $dt^{2},d\phi^{2},dtd\phi.$ This form includes
the magnetic charge which can be removed by a duality
transformation without changing the metric because the sum
$\mathfrak{e}^{2}+\mathfrak{m}^{2}$ remains constant under a
duality transformation of the electromagnetic fields. The
electromagnetic potential is
\begin{equation}\label{emkn}
  \mathbf{A}_{\text{K}}=-\dfrac{\mathfrak{e}r+\mathfrak{m}a\cos\theta}{\varrho^{2}}dt
  +\dfrac{\mathfrak{e}ar\sin^{2}\theta+\mathfrak{m}(r^{2}+a^{2})\cos\theta}{\varrho^{2}}d\phi
\end{equation}
We use subscript K for Kerr-Newman. Comparing with Eq.~\eqref{1}
 we get
\begin{align}
V_{\text{K}} &=1-\dfrac{2M^{\prime}r}{\varrho ^{2}} \label{25} \\
W_{\text{K}} &=-\dfrac{2M^{\prime}ra\sin ^{2}\theta }{\varrho
^{2}}
\label{26} \\ X_{\text{K}} &=(r^{2}+a^{2})\sin ^{2}\theta
+\dfrac{2M^{\prime}ra^{2}\sin ^{4}\theta }{\varrho ^{2}}
\label{27}
\end{align}
In the general spacetime under investigation which is not yet
known to be Kerr-Newman solution we shall choose $r,\theta$
coordinates from Carter's $\rho,z$ coordinates as follows. Let
$r,\theta$ be solution of the following equations with $r\geq
M+\sqrt{M^{2}-\mathfrak{e}^{2}-\mathfrak{m}^{2}-{a}^{2}},$
\begin{equation} \label{rhoandrtheta}
\rho =\sqrt{r^{2}-2M^{\prime}r+a^{2}}\sin \theta, \qquad z=(r-M)\cos \theta
\end{equation}
In the equation for $z$ we use the constant $M$ because $(\partial
\rho/\partial r)=({r^{2}-2M^{\prime }r+a^{2}})^{-1/2}( r-M)
\sin \theta.$ This way
$d\rho^{2}+dz^{2}$ does not have a cross term containing
$drd\theta.$ The expression for $d\rho^{2}+dz^{2}$ is given in
\secref{section6}. $\rho=0$ set which represents the horizon and
the axis is now given by
\begin{equation} \label{rho=0set}
r^{2}-2Mr+a^{2}+\mathfrak{e}^{2}+\mathfrak{m}^{2}=0 \text{, or }
\sin\theta=0
\end{equation}
For convenience we define
\begin{equation}\label{c}
c^{2}=M^{2}-\mathfrak{e}^{2}-\mathfrak{m}^{2}-a^{2}, \qquad c>0
\end{equation}
The restriction on $r$ now becomes $r\geq M+c.$ In general
$(r,\theta)$ coordinate system is defined away from the $\rho=0$
set although the functions $r$ and $\theta$ are defined on this
set. Because of the restriction $r\geq M+c$ the equality is the
only solution of the first equation of Eq.~\eqref{rho=0set}. The
limiting set $r\downarrow M+c$ now contains the horizon and
possibly some parts of the axis while $r> M+c, 0\leq
\theta\leq\pi$ represent the remaining part of $\Sigma^{+}.$ The
portion of the axis in this remaining part of $\Sigma^{+}$ is
later called the part of the axis given by $\theta=0$ or
$\theta=\pi$ \textquotedblleft alone."

\section{Main Idea}

We define some quantities which are crucial for the proof.
\begin{eqnarray}
  2\widetilde{r} &=& r-M+\sqrt{r^{2}-2Mr+\mathfrak{e}^{2}+\mathfrak{m}^{2}+a^{2}}\\
\zeta &=& \widetilde{r}^{2}\varrho^{-2}
\\
  f &=& \widetilde{r}^{2}\sin^{2}\theta
\end{eqnarray}
Significance of these quantities is that they transform the
2-metric
\begin{equation}\label{gkn}
\overline{g}_{\text{K}}=\varrho ^{2}\left(
\dfrac{dr^{2}}{r^{2}-2M^{\prime}r+a^{2}}+d\theta ^{2}\right)
\end{equation}
into the Euclidean 3-metric $\eta_{\text{K}}$ in the spherical
coordinates $\{\widetilde{r},\theta,\phi\}$ as follows
\begin{equation}\label{etaK}
\eta_{\text{K}}=\zeta \overline{g}_{\text{K}}+fd\phi ^{2}=d\widetilde{r}^{2}+\widetilde{r}
^{2}\left( d\theta ^{2}+\sin ^{2}\theta d\phi ^{2}\right)
\end{equation}

Our aim is to show, by exploiting the field equations
(\ref{LV}-\ref{rBD1a}) and reasonable boundary conditions, that
the general 3-metric $\eta$ defined by
\begin{equation} \label{eta}
  \eta=\zeta \overline{g}+fd\phi ^{2}
\end{equation}
where $\zeta$ and $f$ are the same functions of $(r,\theta)$ is
the same Euclidean metric in the coordinates
$\{\widetilde{r},\theta,\phi\}$ where $\widetilde{r}$ is the same
function of $r.$ In the actual process we get $X=X_{\text{K}}$ at
first. Then we show $\Omega=\Omega_{\text{K}}.$ This gives a
single black hole so that the uniqueness proof of Bunting or Mazur applies. Let
\begin{equation} \label{rinout}
r_{\text{out}} =\widetilde{r},
\qquad r_{\text{in}} =(1/2)\left(r-M-\sqrt{r^{2}-2Mr+\mathfrak{e}^{2}+\mathfrak{m}^{2}+a^{2}}\right)
\end{equation}
If we take $\zeta$ and $f$ as $\zeta^{+}$ and $f^{+}$and define
\begin{eqnarray}
\zeta ^{-} &=&r_{\text{in}}^{2}\varrho^{-2} \\
  f^{-} &=&r_{\text{in}}^{2}\sin^{2}\theta
  \end{eqnarray}
then $\eta_{\text{K}}^{-}=\zeta^{-}
\overline{g}_{\text{K}}+f^{-}d\phi
^{2}=dr_{\text{in}}^{2}+r_{\text{in}}^{2}\left( d\theta ^{2}+\sin
^{2}\theta d\phi ^{2}\right)$ is also the Euclidean metric.
Recalling Eq.~\eqref{c} we note that
$r_{\text{in}}=(1/4)c^{2}r_{\text{out}}^{-1}.$ So
\begin{equation} \label{relationpm}
  f^{-}
  =(c^{4}/16)r_{\text{out}}^{-4}f=(16/c^{4})r_{\text{in}}^{4}f,\qquad\qquad
  \zeta ^{-} =(16/c^{4})r_{\text{in}}^{4}\zeta
\end{equation}
We shall use spinor identities for the metric
\begin{equation} \label{chi}
\chi =\sigma^{2}\zeta \overline{g}+Ufd\phi ^{2}
\end{equation}
where $U=U\left(r\right)$ is a solution of a first order ODE with
appropriate boundary conditions to be specified later (see
Eq.~\eqref{qdplus} and \lemref{lemma11.4} below) and
\begin{equation} \label{sigma}
\sigma =\left( X/X_{\text{K}}\right) ^{1/4}>0
\end{equation}
For $r>M+c,$ $\sigma$ is differentiable and positive. This is because for $r>M+c,$
both $X/\sin^{2}\theta,X_{\text{K}}/\sin^{2}\theta$ are positive and regular on the axis.\\
We define $\chi ^{-}=\sigma ^{2}\zeta ^{-}\overline{g}+Uf^{-}d\phi
^{2}.$ Then
\begin{equation}\label{chipm}
\left( c^{4}/16\right) r_{\text{in}}^{-4}\chi ^{-}=\chi ^{+}\equiv \chi
\end{equation}
provided the same function $U$ is used for both the metrics $\chi
^{\pm}.$ The actual functions we shall use are not known to be the
same initially. However Eq.~\eqref{chipm} is useful in
transforming formulas.

$r_{\text{out}}=r_{\text{in}}$ occurs at $r=M\pm c.$ At these
values $r_{\text{out/in}}=\pm c/2=c/2$ neglecting the negative
sign. For a Kerr-Newman solution
 $\eta_{\text{K}}^{-}$ and
$\eta_{\text{K}}^{+}=\eta_{\text{K}}$ match on the boundary sphere
of radius
$r=M+c=M+\sqrt{M^{2}-\mathfrak{e}^{2}-\mathfrak{m}^{2}-{a}^{2}}$
which corresponds to the outer Killing horizon. We have no
business inside the outer Killing horizon.

For the general situation let $\eta^{+}=\eta$ and let $\eta^{-}$
be defined by replacing $f$ and $\zeta$ in Eq.~\eqref{eta} with
$f^{-}$ and $\zeta^{-}$. Asymptotic conditions ensure that
$\eta^{+}$ is asymptotically flat with mass zero and $\eta^{-}$
compactifies the infinity. So if we can show that these metric
have nonnegative scalar curvature and they match smoothly at the
inner boundaries, then positive mass theorem makes them Euclidean.
Since we could not directly show that this scalar curvature is
nonnegative we follow a detour. Keeping the spinorial proof of the
positive mass theorem in mind we construct two spinor identities
that solves difficult parts of the problem.

\section{Computation in $r,$ $\theta$ coordinates} \label{section6}
In general we define $r,\theta$ coordinates using
Eqs~\eqref{rhoandrtheta}. Then we get
\begin{equation} \label{drhodz}
  d\rho^{2}+dz^{2}=\left( r^{2}-2M^{\prime }r+(M^{2}-\mathfrak{e}^{2}-\mathfrak{m}^{2})\sin ^{2}\theta +a^{2}\cos ^{2}\theta \right) \left[
\dfrac{dr^{2}}{r^{2}-2M^{\prime }r+a^{2}}+d\theta ^{2}\right]
\end{equation}

Let
$\Pi=\left(r^{2}-2M^{\prime}r+a^{2}\right)^{-1}dr^{2}+d\theta^{2}.$
We have
\begin{eqnarray}
  \overline{g}&=& \Omega \left( r^{2}-2M^{\prime }r+(M^{2}-\mathfrak{e}^{2}-\mathfrak{m}^{2})\sin ^{2}\theta +a^{2}\cos ^{2}\theta \right)\Pi  \label{ginrtheta}\\
      \overline{g}_{\theta\theta}&=& \Omega
   \left( r^{2}-2M^{\prime }r+(M^{2}-\mathfrak{e}^{2}-\mathfrak{m}^{2})\sin ^{2}\theta +a^{2}\cos ^{2}\theta \right)=|\overline{\nabla \strut}\theta|^{-2}
   \label{gthetatheta} \\
   \overline{g}_{rr} &=&\overline{g}_{\theta\theta}
   \left(r^{2}-2M^{\prime}r+a^{2}\right)^{-1}=   |\overline{\nabla \strut}r|^{-2} \label{grr}
\end{eqnarray}
We note that for $r\geq M+c,$ the expression  $r^{2}-2M^{\prime }r+(M^{2}-\mathfrak{e}^{2}-\mathfrak{m}^{2})\sin ^{2}\theta +a^{2}\cos ^{2}\theta\geq
(M^{2}-\mathfrak{e}^{2}-\mathfrak{m}^{2}-a^{2})\sin^{2}\theta$ is positive away from the axis because we are assuming
$M^{2}>\mathfrak{e}^{2}+\mathfrak{m}^{2}+a^{2}.$ The expression is also positive on the axis for $r> M+c.$
It is useful to remember the formulas
\begin{eqnarray*}
 r^{2}-2M^{\prime }r+(M^{2}-\mathfrak{e}^{2}-\mathfrak{m}^{2})\sin ^{2}\theta +a^{2}\cos ^{2}\theta &=& (r-M-c)(r-M+c)+c^{2}\sin^{2}\theta\\
 r^{2}-2M^{\prime}r+a^{2}&=&(r-M-c)(r-M+c)
\end{eqnarray*}
Only nontrivial Christoffel symbol of $\Pi$ is $\Gamma_{\Pi
rr}^{r}=-\left(r^{2}-2M^{\prime}r+a^{2}\right)^{-1}(r-M).$ All
other Christoffel symbol of $\Pi$ vanish. So using
$\overline{\Delta}_{s\Pi} u=s^{-1}\overline{\Delta}_{\Pi}u$ and
Eq.~\eqref{ginrtheta} we get
\begin{eqnarray}
\overline{\Delta}r&=& (r-M)(r^{2}-2M^{\prime}r+a^{2})^{-1}|\overline{\nabla \strut}r|^{2} \label{lapr}\\
  \overline{\Delta}\theta&=& 0 \label{laptheta}
\end{eqnarray}

We note that
\begin{equation} \label{lapf}
\begin{aligned}
&\overline{\Delta }\ln
\left(f/\sin^{2}\theta)\right)=\overline{\Delta }\ln r_{\text{out}}=0
\\ &\overline{\Delta }\ln f=\overline{\Delta
}\ln(\sin^{2}\theta)=-2\overline{g}^{\theta\theta}\csc^{2}\theta
\end{aligned}
\end{equation}
The first equation follows because in $\mathbb{R}^{2},$ $\ln
r_{\text{out}}$ is a harmonic function. It can also be checked by
explicit calculation using Eq.~\eqref{lapr}. Similarly the second
equation follows because by virtue of Eq.~\eqref{laptheta},
$\overline{\Delta}\ln f=-2\csc^{2}\theta|\overline{\nabla
\strut}\theta|^{2}.$

Eqs.~\eqref{rinout} give
$\dfrac{dr_{\text{out/in}}}{dr}=\pm\dfrac{r_{\text{out/in}}}{\sqrt{r^{2}-2M^{\prime}r+a^{2}}}.$
 For a differentiable function $U=U(r)$ for $r>M+c,$
\begin{equation} \label{dUdr}
\lim\limits_{r \rightarrow (M+c)^{+}}\dfrac{
d \ln U}{d r_{\text{out/in}}}=\pm
\lim\limits_{r \rightarrow (M+c)^{+}}\dfrac{2\sqrt{r^{2}-2M^{\prime}r+a^{2}}}{c}
\dfrac{d \ln U}{d r }
\end{equation}
where $+$ sign of $\pm$ is for $r_{\text{out}}.$ These equations
need some clarification because finally we shall arrange such that
at $r_{\text{out/in}}=c/2,$ $\dfrac{ d \ln U}{d
r_{\text{out}}}=\dfrac{ d \ln U}{d r_{\text{in}}}.$ Thus in the
RHS of Eq.~\eqref{dUdr}, $U=U(r)$ are two different functions
$U^{\pm}$ of $r$ unless $(d\ln U/dr)$ vanishes.

\section{Scalar curvature of the 3-metric $\chi$}
In the following when we use the symbols $X_{\text{K}},W_{\text{K}},V_{\text{K}},
\psi_{\text{K}},\xi_{\text{K}}$ and $\Omega_{\text{K}}$ we mean functions defined
on $\Sigma^{+},r>M+c$ and these functions have the same functions
of the newly defined variables $r,\theta$ on $\Sigma^{+},r>M+c$ as
those of respective functions in the Kerr-Newman solution in the
usual $r,\theta$ coordinates of that solution. Once we establish
the uniqueness these two sets of functions will be the same
object. For example $X_{\text{K}}$ has a factor of
$\sin^{2}\theta$ so it will vanish in the limit as $r\downarrow
M+c,\sin\theta\downarrow 0.$ Now on $\Sigma^{+}$ this limiting set
are the finite parts of the axis between two black holes in
addition to the topmost and bottommost poles (and possibly some
parts of the axis attached to these two poles) but for Kerr-Newman
solution this set consists of only two poles. We cannot expect the
set $r\downarrow M+c$ to have the same horizon-like prorerty for
$X_{\text{K}}$ on $\Sigma^{+}$ as in the Kerr-Newman solution
unless we can show $\Omega_{\text{K}}=\Omega$ on $\Sigma^{+}.$
Also we note that $r=M+c$ set cannot intersect as a curve
transversely the black horizon away from the poles because by
Eq.~\eqref{rho=0set}, $\rho\rightarrow 0$ on the $r=M+c$ set and
then $\rho$ would be $0$ away from the axis and horizon. On the
other hand there is a curve $r=M+c+\epsilon$ for some positive
$\epsilon$ close to the black hole horizon because the $r,\theta$
coordinates are regular there and $\epsilon$ must tend to $0$ as
the horizon is approached with $\sin\theta\ne 0$ on the horizon.
Thus all the black holes will be enclosed inside the limiting set
$r\downarrow M+c.$

We compute the scalar curvature $R_{\chi }$ of $\chi$ defined in
Eq~\eqref{chi}. For convenience we write
\begin{equation*}
  f_{\text{em}}=X^{-1}|\overline{\nabla}\strut
\psi|^{2}+X^{-1}\left|X\overline{\nabla\strut}
\xi+W\overline{\nabla\strut}
\psi\right|^{2}\rho^{-2}-X_{\text{K}}^{-1}|\overline{\nabla}\strut
\psi_{\text{K}}|^{2}-X_{\text{K}}^{-1}\left|X_{\text{K}}\overline{\nabla\strut}
\xi_{\text{K}}+W_{\text{K}}\overline{\nabla\strut}
\psi_{\text{K}}\right|^{2}\rho^{-2}
\end{equation*}
As explained after defining $\sigma$ in Eq.~\eqref{sigma} the expression
$\left\langle \overline{\nabla \strut }\ln \left(
X_{\text{K}}/\sqrt{\rho}\right),\overline{\nabla \strut }\ln \sigma\right\rangle$ is well-defined for $r>M+c.$
\begin{lemma} \label{lemmarchi}
\begin{equation} \label{scalarchi}
\sigma^{2}\zeta R_{\chi }=\dfrac{1}{2}\left\vert \overline{
\nabla \strut }\ln U\right\vert ^{2}+\mathcal{P}-U^{-1}\left( \overline{\Delta }
U+\left\langle \overline{\nabla \strut }U,\overline{\nabla \strut
}\ln f\right\rangle +Q_{\overline{g}}U\right) .
\end{equation}
where $Q_{\overline{g}}$ and $\mathcal{P}$ are as follows.
\begin{eqnarray}
  Q_{\overline{g}} &=& \left(4\left\langle \overline{\nabla \strut }
\ln \left( X_{\text{K}}/\sqrt{\rho }\right) ,\overline{\nabla \strut }\ln
\sigma \right\rangle+f_{\text{em}}\right)^{-} \label{Qg}\\
  \mathcal{P} &=& \left(4\left\langle \overline{\nabla \strut }
\ln \left( X_{\text{K}}/\sqrt{\rho }\right) ,\overline{\nabla \strut }\ln
\sigma \right\rangle+f_{\text{em}}\right)^{+}+8|\overline{\nabla
\strut }\ln \sigma|^{2} \label{P}
\end{eqnarray}
\end{lemma}
\begin{proof}
For a given function $\widetilde{f}$ the scalar curvature
$R_{\gamma }$ of $\gamma =\overline{g}+\widetilde{f}d\phi ^{2}$ is
given by
\begin{equation}\label{scalargamma}
 R_{\gamma }=\overline{R}-\widetilde{f}^{-1}
\overline{\Delta }\widetilde{f}+\dfrac{1}{2}|\overline{\nabla \strut }\ln
\widetilde{f}|^{2}
\end{equation}
 Let $\widetilde{f}=f\zeta^{-1}.$ Then
$\eta=\zeta\gamma.$ So using the conformal transformation formula
\begin{equation}\label{conformula}
  \eta=\Psi ^{4}\gamma, \quad \Psi ^{4}R_{\eta }=R_{\gamma
} -8\Psi^{-1}\Delta _{\gamma }\Psi=R_{\gamma } -8\Delta _{\gamma
}\ln\Psi-8|\nabla \ln\Psi |^{2}_{\gamma}
\end{equation}
and writing the Laplacian $\Delta _{\gamma }$ relative to the
3-metric $\gamma $ in terms of the Laplacian of $\bar{g}$ using
\begin{equation} \label{lap3}
\Delta _{\gamma }u=\overline{\Delta }u+\left( 1/2\right) \left\langle\overline{\nabla
\strut }\ln \widetilde{f},\overline{\nabla \strut }u\right\rangle
\end{equation}
we get $\zeta R_{\eta }=R_{\gamma } -2\zeta^{-1}\overline{\Delta
}\zeta +(3/2)
\zeta ^{-2}|\overline{\nabla \strut }\zeta |^{2}-\left\langle\overline{\nabla \strut }
\ln \widetilde{f},\overline{\nabla \strut }\ln \zeta
\right\rangle.$ Using Eqs.~(\ref{scalargamma},\ref{lx}) we then get
\begin{multline}\label{etaX}
  \zeta R_{\eta }=(1/2)X^{-1}\overline{\Delta }X-(1/2)
\left\langle \overline{\nabla \strut }\ln \rho ,\overline{\nabla \strut }\ln
X\right\rangle +X^{-1}|\overline{\nabla}\strut
\psi|^{2}+X^{-1}\left|X\overline{\nabla\strut}
\xi+W\overline{\nabla\strut}
\psi\right|^{2}\rho^{-2} \\-\widetilde{f}^{-1}\overline{\Delta
}\widetilde{f}+(1/2)|\overline{\nabla \strut }\ln
\widetilde{f}|^{2}-2\zeta ^{-1}\overline{
\Delta }\zeta +(3/2)|\overline{\nabla \strut }\ln \zeta |^{2}-\left\langle
\overline{\nabla \strut }\ln \widetilde{f},\overline{\nabla \strut }\ln
\zeta \right\rangle
\end{multline}
Since for Kerr-Newman this gives
\begin{multline*}
  0=(1/2)X_{\text{K}}^{-1}\overline{\Delta }X_{\text{K}}-(1/2)
\left\langle \overline{\nabla \strut }\ln \rho ,\overline{\nabla \strut }\ln
X_{\text{K}}\right\rangle
+X_{\text{K}}^{-1}|\overline{\nabla}\strut
\psi_{\text{K}}|^{2}+X_{\text{K}}^{-1}\left|X_{\text{K}}\overline{\nabla\strut}
\xi_{\text{K}}+W_{\text{K}}\overline{\nabla\strut}
\psi_{\text{K}}\right|^{2}\rho^{-2}\\ -\widetilde{f}^{-1}\overline{\Delta
}\widetilde{f}+(1/2)|\overline{\nabla \strut }\ln
\widetilde{f}|^{2}-2\zeta ^{-1}\overline{
\Delta }\zeta +(3/2)|\overline{\nabla \strut }\ln \zeta |^{2}-\left\langle
\overline{\nabla \strut }\ln \widetilde{f},\overline{\nabla \strut }\ln
\zeta \right\rangle
\end{multline*}
So using $f_{\text{em}}$ we get
\scriptsize
\begin{align*}
  \zeta R_{\eta }&=(1/2)X^{-1}\overline{\Delta }X-(1/2)
\left\langle \overline{\nabla \strut }\ln \rho ,\overline{\nabla \strut }\ln
X\right\rangle
+f_{\text{em}}-(1/2)X_{\text{K}}^{-1}\overline{\Delta
}X_{\text{K}}+(1/2)
\left\langle \overline{\nabla \strut }\ln \rho ,\overline{\nabla \strut }\ln
X_{\text{K}}\right\rangle \\
  &= (1/2)\overline{\Delta }\ln
X+(1/2)|\overline{\nabla \strut }\ln X|^{2}-(1/2)\overline{\Delta
}\ln X_{\text{K}}-(1/2)|\overline{\nabla
\strut }\ln X_{\text{K}}|^{2}-(1/2)
\left\langle \overline{\nabla \strut }\ln \rho ,\overline{\nabla \strut }\ln
(X/X_{\text{K}})\right\rangle +f_{\text{em}}\\
   &=\dfrac{1}{2}\overline{\Delta }\ln
(X/X_{\text{K}})+(1/2)\left\langle \overline{\nabla \strut
}\ln(XX_{\text{K}}/\rho),\overline{\nabla \strut }\ln
(X/X_{\text{K}})\right\rangle +f_{\text{em}}\\
&=(1/2)\overline{\Delta }\ln
(X/X_{\text{K}})+(1/2)|\overline{\nabla \strut }\ln
(X/X_{\text{K}})|^{2}+(1/2)\left\langle
\overline{\nabla \strut }\ln(X_{\text{K}}^{2}/\rho),\overline{\nabla
\strut }\ln (X/X_{\text{K}})\right\rangle +f_{\text{em}}\\
&=2\overline{\Delta }\ln \sigma+8|\overline{\nabla
\strut }\ln \sigma|^{2}+2\left\langle
\overline{\nabla \strut }\ln(X_{\text{K}}^{2}/\rho),\overline{\nabla
\strut }\ln \sigma\right\rangle +f_{\text{em}}
\end{align*}
\normalsize
where in the last step we used Eq.~\eqref{sigma}. Remembering
$Q_{\overline{g}}$ and $\mathcal{P}$ we get
\begin{equation}\label{retafinal}
  \zeta R_{\eta }=2\overline{\Delta }\ln \sigma+\mathcal{P}-Q_{\overline{g}}
\end{equation}
Next we compute the scalar curvature of
\begin{equation} \label{vartheta}
\vartheta =\sigma ^{2}\eta
\end{equation}
$R_{\vartheta }=\sigma ^{-2}R_{\eta }-4\sigma ^{-3}\Delta _{\eta
}\sigma +2\sigma ^{-4}|\nabla \sigma |_{\eta }^{2}=\sigma
^{-2}R_{\eta }-4\sigma ^{-2}\Delta _{\eta }\ln \sigma -2\sigma
^{-2}|\nabla \ln \sigma |_{\eta }^{2}.$ \ Using Eq.~\eqref{eta}
and the formula Eq.~\eqref{lap3} we have
\begin{equation*}
\Delta _{\eta }\ln \sigma =\overline{\Delta }_{\zeta \overline{g}}\ln \sigma
+\dfrac{1}{2f}\left\langle\overline{\nabla \strut
}f,\overline{\nabla \strut }\ln
\sigma \right\rangle_{\zeta \overline{g}}
\end{equation*}
Thus $\zeta \sigma ^{2}R_{\vartheta }=\zeta R_{\eta
}-4\overline{\Delta }\ln
\sigma -2\left\langle\overline{\nabla \strut }\ln f,\overline{\nabla \strut }\ln \sigma
\right\rangle-2|\nabla \ln \sigma |^{2}.$ Using Eq.~\eqref{retafinal} we get
\begin{equation} \label{Rvartheta}
\zeta \sigma ^{2}R_{\vartheta }=-2\overline{\Delta }\ln \sigma +\mathcal{P}-Q_{\overline{g}}
-2\left\langle \overline{\nabla \strut }\ln \sigma
,\overline{\nabla
\strut }\ln f\right\rangle -2|\overline{\nabla \strut } \ln \sigma |^{2}
\end{equation}
Finally we write $\chi $ as $\vartheta +\varpi d\phi ^{2}.$ $\chi
_{\phi
\phi }=Uf=\sigma ^{2}f+$ $\varpi $ $\Rightarrow $ $U=f^{-1}(\sigma
^{2}f+\varpi).$ We recall that if $\ h=\overline{G}+\varphi d\phi
^{2},$ and $\widehat{h}=
\overline{G}+\widehat{\varphi }d\phi ^{2},$ where $\overline{G}$ is a
2-dimensional metric on the $\phi =$constant surfaces and $\varphi
,\widehat{
\varphi }$ are independent of $\phi ,$ then
\begin{equation} \label{Eqlemma10.1}
R_{h}=R_{\widehat{h}}+\overline{\Delta }_{\overline{G}}\ln
\dfrac{\widehat{
\varphi }}{\varphi }-\dfrac{1}{2}|\overline{\nabla \strut }\ln \varphi |_{
\overline{G}}^{2}+\dfrac{1}{2}|\overline{\nabla \strut }\ln \widehat{\varphi
}|_{\overline{G}}^{2}
\end{equation}
Taking $\overline{G}=\sigma ^{2}\zeta
\overline{g},$ $\varphi =\sigma ^{2}f+\varpi =\chi _{\phi \phi
}$and $\widehat{\varphi }=\sigma ^{2}f=\vartheta _{\phi \phi }$ we
get
\scriptsize
\begin{align*}
  \sigma ^{2}\zeta R_{\chi } & =\sigma ^{2}\zeta\left(R_{\vartheta }+\overline{\Delta }_{\sigma ^{2}\zeta
\overline{g} }\ln \dfrac{\sigma ^{2}f}{\sigma ^{2}f+\varpi
}-(1/2)|\overline{
\nabla \strut }\ln \left( \sigma ^{2}f+\varpi \right) |_{\overline{G}}^{2}+
(1/2)|\overline{\nabla \strut }\ln \left( \sigma ^{2}f\right) |_{
\overline{G}}^{2}\right) \\
   & =\sigma ^{2}\zeta\left(R_{\vartheta }+\sigma ^{-2}\zeta ^{-1}\overline{\Delta }\ln
\dfrac{\sigma ^{2}f}{\sigma ^{2}f+\varpi }-(1/2)\sigma
^{-2}\zeta ^{-1}|\overline{
\nabla \strut }\ln \left( \sigma ^{2}f+\varpi \right) |^{2}+(1/2)
\sigma ^{-2}\zeta ^{-1}|\overline{\nabla \strut }\ln \left( \sigma
^{2}f\right) |^{2}\right)\\
   & =\sigma ^{2}\zeta R_{\vartheta }+\overline{\Delta }\ln
(\sigma ^{2}U^{-1})-(1/2)|\overline{
\nabla \strut }\ln(Uf) |^{2}+(1/2)
|\overline{\nabla \strut }\ln \left( \sigma ^{2}f\right) |^{2}\\
   & =\sigma ^{2}\zeta R_{\vartheta }+\overline{\Delta }\ln (\sigma ^{2}U^{-1})-(1/2)|\overline{
\nabla \strut }\ln U |^{2}+2
|\overline{\nabla \strut }\ln \sigma|^{2}-\left\langle\overline{
\nabla \strut }\ln U,\overline{
\nabla \strut }\ln f\right\rangle+2\left\langle\overline{
\nabla \strut }\ln \sigma,\overline{
\nabla \strut }\ln f\right\rangle\\
& =-2\overline{\Delta }\ln \sigma
+\mathcal{P}-Q_{\overline{g}}+\overline{\Delta }\ln (\sigma
^{2}U^{-1})-(1/2)|\overline{
\nabla \strut }\ln U |^{2}-\left\langle\overline{
\nabla \strut }\ln U,\overline{
\nabla \strut }\ln f\right\rangle
\end{align*}
\normalsize
where in the last step we used Eq.~\eqref{Rvartheta}. Thus we get
Eq.~\eqref{scalarchi}.
\end{proof}

Similarly we find the scalar curvature of
$\chi^{-}=\sigma^{2}\zeta^{-}
\overline{g}+Uf^{-}d\phi ^{2}$ to be
\begin{equation}\label{scalarchiminus}
R_{\chi^{-}}=(\zeta^{-})^{-1}\sigma^{-2}\left(\mathcal{P}-Q_{\overline{g}}-U^{-1}\overline{\Delta
}U +\dfrac{1}{2}|\overline{
\nabla \strut }\ln U |^{2}-\left\langle
\overline{\nabla \strut }\ln U,\overline{\nabla \strut }\ln
f^{-}\right\rangle\right)
\end{equation}
We can also derive this formula by conformal transformation
$\chi^{-}=(16/c^{4})r_{\text{in}}^{4}\chi$ (see Eq.~\eqref{chipm})
and the fact that $\ln r_{\text{in}}$ is a harmonic function in
the 2-metric $\overline{g}.$

\section{Finding a Spinor}

Let $r _{0}$ be a constant. Let $\oiint\limits_{r =r_{0},\eta}$
represents the surface integral on the $r=r_{0}$ surface relative
to the 2-metric induced from $\eta $ and $\iint\limits_{r =r_{0}}$
represents ordinary double integral. All integrations are done on
subsets of $\Sigma ^{+}\cup \partial
\Sigma ^{+}$ unless indicated otherwise.

We need a $SU(2)$-spinor $\Theta_{\vartheta}$ on $\Sigma$ with the
following properties.
\begin{eqnarray}
  &D_{\vartheta}\Theta_{\vartheta}=0 \label{dirac1}\\
  &||\Theta_{\vartheta}||=1+O(r^{-1}) \as r \rightarrow \infty \label{thetasym}\\
  &\dfrac{\partial ||\Theta_{\vartheta} ||^{2}}{\partial r}=o(r^{-1}) \as r \rightarrow \infty \label{dthetasym}
  \\
  &\Theta_{\vartheta} \text{ is independent of } \phi \label{nophi}\\
&\text{As } r_{0}\downarrow M+c \text{ on the horizon }
\int\limits_{r=r_{0}}\sqrt{r^{2}-2M^{\prime}r+a^{2}}
\dfrac{\partial||\Theta_{\vartheta} ||^{2}}{\partial
r}d\theta =0 \label{dthetahor}
\end{eqnarray}
Here $D_{\vartheta}$ is the Dirac operator of the metric
$\vartheta.$ Such a spinor exists. On the double the 3-metric
$\widehat{g}$ has nonnegative scalar curvature, and it is
asymptotically flat. So we can use Bartnik's proof for the
existence of a spinor $\Theta_{\widehat{g}}$ harmonic relative to
$\widehat{g}.$ Because of the axisymmtery we can choose
$\Theta_{\widehat{g}}$ to be independent of $\phi.$
$\Theta_{\vartheta}$ can be obtained $\Theta_{\widehat{g}}$ by
what we called a 2+1 conformal transformation. It is explained
below.
 We have outlined the
proof of the following lemma in the appendix.
\begin{lemma} \label{lemma7}
Let $\overline{G}=\overline{G}_{11}(d(x^{1})^{2}+d(x^{2})^{2}),$
$g_{1}=\overline{G}+f_{1}d\phi^{2}$ and
$g_{2}=\overline{G}+qf_{1}d\phi^{2}.$ All functions and metrics
are independent of $\phi.$ If $\Theta $ is a spinor satisfying the
Dirac equation $D_{g_{1}}\Theta =0$ and $\Theta $ is independent
of $\phi,$ then
\begin{equation*}
D_{g_{2}}\left( q^{-\dfrac{3}{8}}\Theta \right) =0
\end{equation*}
\end{lemma}

We also have the conformal transformation formula. Let $\xi_{\psi
^{-2}\widehat{\chi}}$ be a fixed spinor satisfying Dirac equation
relative to the metric $\psi ^{-2}\widehat{\chi} $. Then the
spinor $\xi _{\widehat{\chi }}=\psi ^{-1}\xi_{\psi
^{-2}\widehat{\chi}}$ satisfies Dirac equation relative to the
conformal metric $\widehat{\chi} $ (Lichnerowicz \cite{L},
Branson, T., Kosmann-Schwarzbach
\cite{BK}).

To find the spinor $\Theta _{\vartheta }$ from $\Theta
_{\widehat{g}}$ we take
$g_{1}=\widehat{g}=\overline{g}+Xd\phi^{2}$ and
$g_{2}=\sigma^{-2}\zeta^{-1}\vartheta =\overline{g}+\zeta^{-1}f
d\phi ^{2}$ in \lemref{lemma7}. That is we put $f_{1}=X$ and
$q=f\zeta^{-1} X^{-1}.$ Then $D_{\sigma ^{-2}\zeta ^{-1}\vartheta
}\left( \left( f\zeta^{-1}X^{-1}\right) ^{-3/8}\Theta
_{\widehat{g}}\right) =0.$ So the spinor
\begin{equation} \label{spvartheta}
\Theta _{\vartheta }=\sigma ^{-1}\zeta ^{-1/2}\left(
f\zeta^{-1}X^{-1}\right) ^{-3/8}\Theta _{\widehat{g}}
\end{equation}
satisfies $D_{\vartheta }\Theta _{\vartheta }=0.$
We note that
\begin{equation} \label{varandg}
  ||\Theta _{\vartheta }||^{2}=\sigma^{-2}\zeta^{-1/4} X^{3/4}f^{-3/4}||\Theta
_{\widehat{g}}||^{2}
\end{equation}
Now $8\pi M=-\oiint\limits_{\lambda _{0}\rightarrow \infty
,\overline{g}}\left\langle \nabla \strut
_{\widehat{g}}||\Theta_{\widehat{g}} ||^{2},n_{\widehat{g
}}\right\rangle _{\widehat{g }}$ so that (because of the
asymptotic regularity in the existence proof of the spinor)
\begin{equation}\label{Thetasqasym}
  ||\Theta_{\widehat{g}}||^{2}=1-2M/r+O(r^{-2})
\end{equation}
which by virtue of the asymptotic conditions on $\sigma,f,\zeta,X$
gives
\begin{equation}\label{Thetasqasym}
  ||\Theta_{\vartheta}||^{2}=1+O(r^{-2})
\end{equation}
Thus $\Theta_{\vartheta}$ satisfies properties
\ref{dirac1}-\ref{nophi}. To see Eq.~\eqref{dthetahor} we note that
\\
\scriptsize
$\int\limits_{r=r_{0}}\sqrt{r^{2}-2M^{\prime}r+a^{2}}
\dfrac{\partial||\Theta_{\vartheta} ||^{2}}{\partial
r}d\theta=-\oint\limits_{r_{0},\overline{\vartheta} }\left\langle
\overline{
\nabla \strut }||\Theta_{\vartheta} ||^{2},n_{\overline{\vartheta}}
\right\rangle_{\overline{\vartheta}}$ \normalsize
where $n_{\overline{\vartheta}}$ is the
unit normal form on the $r=\text{constant}$ loop on a
$\phi=\text{constant}$ surface with the normal vector pointing
towards decreasing $r.$ $\Theta_{\widehat{g}}$ is a harmonic
spinor on the double $\Sigma^{+}\cup \Sigma^{-}\cup \partial
\Sigma^{+}$ relative to the metric $\widehat{g}$ producing the
same contribution at each end for
$\lim\limits_{r_{0}\uparrow\infty}\oiint\limits_{r_{0},\widehat{g}}
\left\langle
\overline{
\nabla \strut }||\Theta_{\widehat{g}} ||^{2},n_{\widehat{g}}
\right\rangle_{\widehat{g}}.$ Thus on the horizon
$\oiint\limits_{\widehat{g} }\left\langle
\overline{
\nabla \strut }||\Theta _{\widehat{g}
}||^{2},n_{\widehat{g}}\right\rangle_{\widehat{g}}=0$ because of
the symmetry across the totally geodesic boundary where
$n_{\widehat{g}}$ is the unit normal form on the relevant surface.
Since the derivative of $\sigma^{-2}\zeta^{-1/4} X^{3/4}f^{-3/4}$
is regular on the horizon we get Eq.~\eqref{dthetahor}.

Now we find a spinor satisfying $D_{\chi }\Theta _{\chi }=0.$
Recalling Eqs.~(\ref{vartheta},\ref{eta},\ref{chi}) for $\chi $
and $\vartheta $ we apply \lemref{lemma7} again with
$g_{1}=\vartheta ,$ $f_{1}=f\sigma ^{2},$ $g_{2}=\chi $ and
$q=U\sigma ^{-2}.$ Thus we take
\begin{equation} \label{spchi}
\Theta _{\chi }=\sigma ^{\dfrac{3}{4}}U^{-\dfrac{3}{8}}\Theta _{\vartheta }
\end{equation}
For the harmonic spinor $\Theta _{\chi }$ one has the identity,
\begin{equation} \label{lich2}
2\Delta _{\chi }||\Theta _{\chi }||^{2}=R_{\chi }||\Theta _{\chi
}||^{2}+4||\nabla _{\chi }\Theta _{\chi }||^{2}
\end{equation}
Using the expression for $R_{\chi }$ from Eq.~\eqref{scalarchi} we
get
\scriptsize
\begin{equation} \label{lich3}
2\Delta _{\chi }||\Theta _{\chi }||^{2}=\left((1/2)\left\langle
\overline{\nabla
\strut }\ln U,\overline{\nabla \strut }\ln U\right\rangle+\mathcal{P}-U^{-1}\left( \overline{\Delta }
U+\left\langle \overline{\nabla \strut }U,\overline{\nabla \strut
}\ln f\right\rangle +Q_{\overline{g}}U\right)\right)\sigma
^{-2}\zeta^{-1}||\Theta _{\chi }||^{2}+4||\nabla _{\chi }\Theta
_{\chi }||^{2}
\end{equation}
\normalsize
For complex $U,$ $|\overline{\nabla
\strut }\ln U|^{2}$ is not nonnegative definite. So we have replaced it in Eq.~\eqref{scalarchi} by $\left\langle
\overline{\nabla
\strut }\ln U,\overline{\nabla \strut }\ln U\right\rangle$ to remove confusion. Our next aim
is to write the above identity using 2-dimensional
Laplacian because that way we can easily tackle integration if $U$
becomes complex.

\section{Two spinor identities on $\Sigma^{\pm}$} \label{sec11}
Let $L=\left(r,\theta \right)$
\begin{equation} \label{L}
|U|^{-\dfrac{3}{4}}/||\Theta_{\chi} ||^{2}=L
\end{equation}
$L$ is not defined on the possible zero set of $||\Theta _{\chi
}||$ in case the known spinor $\Theta_{\widehat{g}} $ can vanish.
By Eq.~\eqref {spchi},
$\sigma^{3/2}L||\Theta_{\vartheta}||^{2}=1.$ We need to introduce
$L$ on $\Sigma^{\pm}$ only for $L_{ave}\leq 4/3.$ This will be
clear later from Eqs.~(\ref{cqdp},\ref{cqdm}).
$L_{\text{ave}}=L_{\text{ave}}(r), r> M+c$ is defined as follows.
\begin{equation} \label{lave}
L_{ave}(r_{0})\int\limits_{r=r_{0}}\sigma^{3/2}||\Theta_{\vartheta}
||^{2}\sin^{2}\theta d\theta=\int\limits_{r=r_{0}}\sigma^{3/2}L
||\Theta_{\vartheta} ||^{2}\sin^{2}\theta
d\theta=\int\limits_{r=r_{0}}\sin^{2}\theta d\theta
\end{equation}
We shall use $L$ either in the form $L||\Theta_{\vartheta}||^{2}$
or in $L_{\text{ave}}.$
 We removed the $\theta$ dependence from $L=L\left(
r,\theta
\right)$ by averaging $L$ on the $r=\text{constant}$ loops on a
$\phi=\text{constant}$ surface. For future reference we note that
$L$ and $L_{\text{ave}}
\rightarrow 1$ as $r \rightarrow \infty.$ This follows from
Eq.~\eqref{thetasym}. We have introduced $L$ to present some
equations in a compact form. We can possibly take the mystery out
of $L$ by examining the following cumbersome expression for $L.$
\begin{equation*}
  L=X_{\text{K}}^{-1/8}\zeta^{1/4}X^{-5/8}f^{3/4}||\Theta_{\widehat{g}}||^{-2}
\end{equation*}
Thus as $\sin\theta \rightarrow 0$ but $r> M+c,$
$L=O(1)\zeta^{1/4}||\Theta_{\widehat{g}}||^{-2}.$ As $\sin\theta
\rightarrow 0$ and $r\downarrow M+c,$
$L=O((\sin\theta)^{1/4})\zeta^{1/4}||\Theta_{\widehat{g}}||^{-2}.$
Thus $L$ tends to a finite limit as $r\downarrow M+c$ on the
horizon whenever the spinor $\Theta_{\widehat{g}}\ne 0$ on the
horizon. Since $\Theta_{\widehat{g}}$ cannot identically vanish on
the totally geodesic surface representing the horizon,
Eq.~\eqref{lave} shows that $L_{\text{ave}}$ is bounded as
$r\downarrow M+c.$ However as stated above we are interested only
for $L_{ave}\leq 4/3.$

We denote a $\phi=\text{constant}$ surface in $\Sigma^{+}$ by
$\Sigma_{2}^{+}.$ Let $U=|U|e^{i\omega},$ $\omega$ being real. We
assume that $U,\omega$ are functions of $r$ only.
\begin{lemma} \label{lemma11.1}
On $(\Sigma_{2}^{+},\overline{\chi})$ for $U=U(r)$ wherever
$U=U(r)$ is twice differentiable,
\scriptsize
\begin{multline}\label{id}
  \overline{\nabla \strut }_{\overline{\chi
}}\left(\sin^{2}\theta\left( 2\overline{
\nabla \strut }||\Theta _{\chi }||^{2}+
2||\Theta _{\chi }||^{2}\overline{
\nabla \strut }\ln r_{\text{out}}-2|U|^{-\dfrac{3}{4}}\overline{\nabla \strut }\ln
r_{\text{out}}+||\Theta _{\chi }||^{2}\overline{\nabla
\strut }\ln U\right)-(||\Theta _{\chi }||^{2}-1)\overline{
\nabla \strut }\sin^{2}\theta\right) =\\
-||\Theta _{\chi }||^{2}\overline{\Delta
}_{\overline{\chi}}\sin^{2}\theta +\overline{\Delta
}_{\overline{\chi}}\sin^{2}\theta + 4\sin^{2}\theta||\nabla _{\chi
}\Theta _{\chi }||^{2} -
\dfrac{3}{4}iL\left\langle
\overline{\nabla \strut }\omega ,\overline{
\nabla \strut }\ln f\right\rangle_{\overline{\chi}} ||\Theta _{\chi }||^{2}\sin^{2}\theta+\\
\sin^{2}\theta\left( \mathcal{P}-\dfrac{1}{2}\left\langle
\overline{\nabla \strut }\ln U,\overline{\nabla \strut }\ln
U\right\rangle-\dfrac{1}{4}(4-3L)\left\langle
\overline{\nabla \strut }\ln U,\overline{\nabla \strut }\ln
f\right\rangle -Q_{\overline{g}}\right)\zeta ^{-1}||\Theta _{\chi
}||^{2}
\end{multline}
\normalsize
\end{lemma}
\begin{proof}
Writing the 3-Laplacian $\Delta _{\chi }$ relative to $\chi
=\sigma^{2}\zeta
\overline{g}+Ufd\phi ^{2}$ in terms of the Laplacian of the 2-metric
$\overline{\chi}=\sigma^{2}\zeta \overline{g}$ we get
\scriptsize
$2\Delta_{\chi}||\Theta_{\chi} ||^{2}=2\overline{\Delta
}_{\overline{\chi }}||\Theta_{\chi} ||^{2}+\left\langle
\overline{\nabla \strut }\ln \left( Uf\right) ,
\overline{\nabla \strut }||\Theta_{\chi}  ||^{2}\right\rangle _{\overline{
\chi }}.$\normalsize \,\, We also have \scriptsize
$U^{-1}||\Theta _{\chi }||^{2}\overline{\Delta }_{\overline{
\chi }}U=\overline{\nabla \strut }_{\overline{\chi }}\left( ||\Theta _{\chi
}||^{2}U^{-1}\overline{\nabla \strut }U\right) -\left\langle
\overline{
\nabla \strut }||\Theta _{\chi }||^{2},\overline{\nabla \strut }\ln
U\right\rangle _{\overline{\chi }}+\left\langle \overline{\nabla
\strut }\ln U,\overline{\nabla \strut }\ln U\right\rangle
_{\overline{\chi }}||\Theta _{\chi }||^{2}.$\normalsize \,\, So
Eq.~\eqref{lich3} gives \scriptsize $2\overline{\Delta
}_{\overline{\chi }}||\Theta _{\chi }||^{2}+\left\langle
\overline{\nabla \strut }\ln f,\overline{\nabla \strut }
||\Theta _{\chi }||^{2}\right\rangle
_{\overline{\chi}}+\overline{\nabla \strut }_{\overline{\chi
}}\left( ||\Theta _{\chi }||^{2}U^{-1}\overline{\nabla \strut
}U\right)=\\
\left( \mathcal{P}-(1/2)\left\langle
\overline{\nabla \strut }\ln U,\overline{\nabla \strut }\ln U\right\rangle
-\left\langle \overline{\nabla \strut }\ln U,\overline{\nabla
\strut }\ln f\right\rangle -Q_{\overline{g}}\right)\sigma^{-2}\zeta ^{-1}||\Theta _{\chi }||^{2}+4||\nabla _{\chi}\Theta _{\chi
}||^{2}\Rightarrow $\\ $2\overline{\Delta }_{\overline{\chi
}}||\Theta _{\chi }||^{2}+\overline{
\nabla \strut }_{\overline{\chi }}\left( ||\Theta _{\chi }||^{2}\overline{
\nabla \strut }\ln f-2|U|^{-3/4}\overline{\nabla \strut }\ln
r_{\text{out}}+||\Theta _{\chi }||^{2}\overline{\nabla \strut }\ln
U\right) = 4||\nabla _{\chi }\Theta _{\chi }||^{2}+||\Theta _{\chi
}||^{2}\sigma^{-2}\zeta ^{-1}\overline{\Delta }\ln f+\\
(3/2)|U|^{-3/4}
\left\langle \overline{\nabla \strut }\ln |U|,\overline{
\nabla \strut }\ln r_{\text{out}}\right\rangle
+\left( \mathcal{P}-(1/2)\left\langle
\overline{\nabla \strut }\ln U,\overline{\nabla \strut }\ln U\right\rangle
-\left\langle \overline{\nabla \strut }\ln U,\overline{\nabla
\strut }\ln f\right\rangle -Q_{\overline{g}}\right)\sigma^{-2}\zeta ^{-1}||\Theta _{\chi
}||^{2}.$\normalsize \,\, Now we use Eq.~\eqref{lapf} to get
\scriptsize
\begin{multline}\label{Hsides}
   \overline{\nabla \strut
}_{\overline{\chi }}\left( 2\overline{
\nabla \strut }||\Theta _{\chi }||^{2}+
||\Theta _{\chi }||^{2}\overline{
\nabla \strut }\ln f-2|U|^{-3/4}\overline{\nabla \strut }\ln
r_{\text{out}}+||\Theta _{\chi }||^{2}\overline{\nabla \strut }\ln
U\right) =4||\nabla _{\chi }\Theta _{\chi }||^{2}- 2||\Theta
_{\chi }||^{2}\sigma^{-2}\zeta ^{-1}\overline{g}^{\theta\theta}
\csc^{2}\theta+\\(3/2)|U|^{-3/4}\left\langle \overline{
\nabla \strut }\ln |U|,\overline{\nabla \strut }\ln r_{\text{out}}\right\rangle_{\overline{\chi}}
+\left(
\mathcal{P}-(1/2)\left\langle
\overline{\nabla \strut }\ln U,\overline{\nabla \strut }\ln U\right\rangle
-\left\langle \overline{\nabla \strut }\ln U,\overline{\nabla
\strut }\ln f\right\rangle -Q_{\overline{g}}\right)\sigma^{-2}\zeta ^{-1}||\Theta _{\chi
}||^{2}
\end{multline}
\normalsize
For $U=U(r),$ $2\left\langle \overline{\nabla \strut }\ln |U|,
\scriptsize
\overline{\nabla \strut }\ln r_{\text{out}}\right\rangle
=\left\langle \overline{\nabla \strut }\ln |U|,
\overline{\nabla \strut }\ln f\right\rangle =\left\langle \overline{\nabla
\strut }\ln U,\overline{\nabla \strut }\ln f\right\rangle -i\left\langle
\overline{\nabla \strut }\omega ,\overline{\nabla \strut }\ln f\right\rangle
.$\normalsize \,\, So RHS of Eq.~\eqref{Hsides} simplifies to
\begin{multline*}
4||\nabla _{\chi }\Theta _{\chi }||^{2}-2 ||\Theta _{\chi
}||^{2}\sigma^{-2}\zeta ^{-1}\overline{g}^{\theta\theta}
\csc^{2}\theta
-(3/4)iL\left\langle
\overline{\nabla \strut }\omega ,\overline{
\nabla \strut }\ln f\right\rangle_{\overline{\chi}} ||\Theta _{\chi }||^{2}+\\ \left( \mathcal{P}-(1/2)\left\langle
\overline{\nabla \strut }\ln U,\overline{\nabla \strut }\ln U\right\rangle
-\dfrac{1}{4}(4-3L)\left\langle \overline{\nabla \strut }\ln
U,\overline{\nabla \strut }\ln f\right\rangle
-Q_{\overline{g}}\right)\sigma^{-2}\zeta ^{-1}||\Theta _{\chi
}||^{2}
\end{multline*}
Now we multiply both sides of Eq.~\eqref{Hsides} by
$\sin^{2}\theta.$ LHS becomes (since $U, r_{\text{out}}$ are
functions of $r$ only), \\
\scriptsize
$\overline{\nabla \strut }_{\overline{\chi
}}\left(\sin^{2}\theta\left( 2\overline{
\nabla \strut }||\Theta _{\chi }||^{2}+
||\Theta _{\chi }||^{2}\overline{
\nabla \strut }\ln f-2|U|^{-3/4}\overline{\nabla \strut }\ln
r_{\text{out}}+||\Theta _{\chi }||^{2}\overline{\nabla
\strut }\ln U\right)\right)- \\ 2\left\langle\overline{\nabla \strut
}\sin^{2}\theta,\overline{
\nabla \strut }||\Theta _{\chi }||^{2}\right\rangle_{\overline{\chi }}
-\left\langle\overline{\nabla \strut }\sin^{2}\theta,\overline{
\nabla \strut }\ln(\sin^{2}\theta)\right\rangle_{\overline{\chi }}||\Theta _{\chi }||^{2}
 =$ \\
$\overline{\nabla \strut }_{\overline{\chi
}}\left(\sin^{2}\theta\left( 2\overline{
\nabla \strut }||\Theta _{\chi }||^{2}+
||\Theta _{\chi }||^{2}\overline{
\nabla \strut }\ln f-2|U|^{-3/4}\overline{\nabla \strut }\ln
r_{\text{out}}+||\Theta _{\chi }||^{2}\overline{\nabla
\strut }\ln U\right)\right)- 2\overline{\nabla \strut
}_{\overline{\chi }}
\left(||\Theta _{\chi }||^{2}\overline{\nabla \strut }_{\overline{\chi
}}\sin^{2}\theta
\right)+2||\Theta _{\chi }||^{2}\overline{\Delta \strut }_{\overline{\chi
}}\sin^{2}\theta \\ -4\sigma^{-2}\zeta
^{-1}\overline{g}^{\theta\theta}||\Theta _{\chi
}||^{2}\cos^{2}\theta
 =$\\
$\overline{\nabla \strut }_{\overline{\chi
}}\left(\sin^{2}\theta\left( 2\overline{
\nabla \strut }||\Theta _{\chi }||^{2}+
2||\Theta _{\chi }||^{2}\overline{
\nabla \strut }\ln r_{\text{out}}+2||\Theta _{\chi }||^{2}\overline{
\nabla \strut }\ln \sin\theta-2|U|^{-3/4}\overline{\nabla \strut }\ln
r_{\text{out}}+||\Theta _{\chi }||^{2}\overline{\nabla
\strut }\ln U\right)\right)- 2\overline{\nabla \strut
}_{\overline{\chi }}
\left(||\Theta _{\chi }||^{2}\overline{\nabla \strut }_{\overline{\chi
}}\sin^{2}\theta
\right)\\+2||\Theta _{\chi }||^{2}\overline{\Delta \strut }_{\overline{\chi }}\sin^{2}\theta
-4\sigma^{-2}\zeta ^{-1}\overline{g}^{\theta\theta}||\Theta _{\chi
}||^{2}\cos^{2}\theta
  =$ \\
  $\overline{\nabla \strut }_{\overline{\chi
}}\left(\sin^{2}\theta\left( 2\overline{
\nabla \strut }||\Theta _{\chi }||^{2}+
2||\Theta _{\chi }||^{2}\overline{
\nabla \strut }\ln r_{\text{out}}-2|U|^{-3/4}\overline{\nabla \strut }\ln
r_{\text{out}}+||\Theta _{\chi }||^{2}\overline{\nabla
\strut }\ln U\right)-||\Theta _{\chi }||^{2}\overline{
\nabla \strut }\sin^{2}\theta\right)+2||\Theta _{\chi }||^{2}\overline{\Delta \strut }_{\overline{\chi
}}\sin^{2}\theta \\-4\sigma^{-2}\zeta
^{-1}\overline{g}^{\theta\theta}||\Theta _{\chi
}||^{2}\cos^{2}\theta
 =$ \\
  $\overline{\nabla \strut }_{\overline{\chi
}}\left(\sin^{2}\theta\left( 2\overline{
\nabla \strut }||\Theta _{\chi }||^{2}+
2||\Theta _{\chi }||^{2}\overline{
\nabla \strut }\ln r_{\text{out}}-2|U|^{-3/4}\overline{\nabla \strut }\ln
r_{\text{out}}+||\Theta _{\chi }||^{2}\overline{\nabla
\strut }\ln U\right)-||\Theta _{\chi }||^{2}\overline{
\nabla \strut }\sin^{2}\theta\right)
-4\sigma^{-2}\zeta ^{-1}\overline{g}^{\theta\theta}||\Theta _{\chi
}||^{2}\sin^{2}\theta $\\
 \normalsize
where we used
\begin{equation}\label{lapsin2}
  \overline{\Delta }\sin^{2}\theta=2\overline{g}^{\theta\theta}\cos(2\theta)
\end{equation}
Thus Eq.~\eqref{Hsides} becomes
\scriptsize
\begin{multline}\label{2}
  \overline{\nabla \strut }_{\overline{\chi
}}\left(\sin^{2}\theta\left( 2\overline{
\nabla \strut }||\Theta _{\chi }||^{2}+
2||\Theta _{\chi }||^{2}\overline{
\nabla \strut }\ln r_{\text{out}}-2|U|^{-\dfrac{3}{4}}\overline{\nabla \strut }\ln
r_{\text{out}}+||\Theta _{\chi }||^{2}\overline{\nabla
\strut }\ln U\right)-||\Theta _{\chi }||^{2}\overline{
\nabla \strut }\sin^{2}\theta\right) =\\
-2\sigma^{-2}\zeta ^{-1}\overline{g}^{\theta\theta}||\Theta _{\chi
}||^{2}\cos(2\theta)+ 4\sin^{2}\theta||\nabla _{\chi }\Theta
_{\chi }||^{2} -
 (3/4)iL\left\langle
\overline{\nabla \strut }\omega ,\overline{
\nabla \strut }\ln f\right\rangle_{\overline{\chi}} ||\Theta _{\chi }||^{2}\sin^{2}\theta
\\+\sin^{2}\theta\left( \mathcal{P}-\dfrac{1}{2}\left\langle
\overline{\nabla \strut }\ln U,\overline{\nabla \strut }\ln U\right\rangle
-(1/4)(4-3L)\left\langle \overline{\nabla \strut }\ln
U,\overline{\nabla \strut }\ln f\right\rangle
-Q_{\overline{g}}\right)\sigma^{-2}\zeta ^{-1}||\Theta _{\chi
}||^{2}
\end{multline}
\normalsize
Hence using Eq.~\eqref{lapsin2} again and adding $\overline{\Delta
}_{\overline{\chi}}\sin^{2}\theta$ to both sides we get
Eq.~\eqref{id}.
\end{proof}

On $(\Sigma^{-},\overline{\chi^{-}})$ we get a similar identity as
Eq.~\eqref{id} using the spinor
$\Theta_{\chi^{-}}=\sigma^{3/4}U^{-3/8}\Theta_{\vartheta^{-}}$
(compare Eq.~\eqref{spchi}) where
$\vartheta^{-}=\sigma^{2}\eta^{-}=\sigma^{2}\zeta^{-}\overline{g}+\sigma^{2}f^{-}d\phi^{2}$
and $\Theta_{\vartheta^{-}}$ we define replacing $\zeta,f$ by
$\zeta^{-},f^{-}$ in Eq.~\eqref{spvartheta}. Then by
Eq.~\eqref{relationpm}, $\Theta_{\chi^{-}}
=(c^{2}/4)r_{\text{in}}^{-2}\sigma^{3/4}U^{-3/8}\Theta_{\vartheta}
=(c^{2}/4)r_{\text{in}}^{-2}\Theta_{\chi}.$ This is
$\Theta_{\chi}$ on $\Sigma^{-}$ and it may differ from
$\Theta_{\chi}$ on $\Sigma^{+}$ because in general $U(r)$ are
different functions on $\Sigma^{\pm}.$ However to keep notation
simple we use $U$ for $U^{\pm}.$ We also recall
$\overline{\chi^{-}} =(16/c^{4})r_{\text{in}}^{4}\overline{\chi}.$
The singular factor $r_{\text{in}}^{-2}$ in $\Theta_{\chi^{-}}$
makes it difficult to manipulate. So we shall write the identity
for $\Sigma^{-}$ using $\Theta_{\chi}$ on $\Sigma^{-}.$ This is
done below using several transformation formulas which are
straightforward to check.
\scriptsize
\begin{eqnarray}
||\nabla
_{\chi^{-}}(r_{\text{in}}^{-2}\Theta_{\chi})||^{2}&=&4r_{\text{in}}^{-4}|\nabla
\ln
r_{\text{in}}|^{2}_{\chi^{-}}||\Theta_{\chi}||^{2}+r_{\text{in}}^{-4}||\nabla
_{\chi_{-}}\Theta_{\chi}||^{2}-2r_{\text{in}}^{-4}\left\langle
\nabla \ln
r_{\text{in}},\nabla ||\Theta_{\chi}||^{2}\right\rangle_{\chi^{-}}
\label{done} \\
(16/c^{4})||\nabla
_{\chi_{-}}\Theta_{\chi}||^{2}&=&r_{\text{in}}^{-4}||\nabla
_{\chi}
\Theta_{\chi}||^{2}- r_{\text{in}}^{-4}\left\langle \nabla
\ln r_{\text{in}},\nabla
||\Theta_{\chi}||^{2}\right\rangle_{\chi}+2r_{\text{in}}^{-4}|\nabla
\ln r_{\text{in}}|^{2}_{\chi}||\Theta_{\chi}||^{2} \label{101} \\
 \Delta_{\chi^{-}}\ln r_{\text{in}}&=&(1/2)\left\langle
\overline{\nabla \strut }\ln \left( Uf^{-}\right) ,
\overline{\nabla \strut }\ln
r_{\text{in}}\right\rangle _{\chi^{-} } \label{3laplnr}
\end{eqnarray}
\normalsize
As usual when not specified explicitly norms and inner products
for gradients of functions are w.r.t. ${\overline{g}}.$ In order
to tackle the contribution from the term \scriptsize
$||\Theta_{\chi}||^{2}\left\langle
\overline{\nabla \strut }\ln U,
\overline{\nabla \strut }\ln
r_{\text{in}}\right\rangle _{\chi^{-} }$ \normalsize coming from
Eq.~\eqref{3laplnr} we shall need the following lemma. The lemma
is proved in the appendix.
\begin{lemma} \label{newlemma100}
For the metric $\gammaup=\sigma
^{2}\zeta\overline{g}+|U|^{-4}Ufd\phi^{2}$ and spinor
$\Theta_{\gammaup}=|U|^{1/2}\Theta_{\chi},$
\begin{equation}\label{neweq2d}
  ||\nabla_{\chi}\Theta_{\chi}||^{2}-(1/2)\sigma^{-2}\zeta^{-1}\left\langle
\overline{\nabla \strut} \ln U,\overline{\nabla \strut} \ln
f\right\rangle||\Theta_{\chi}||^{2}
=iI+|U|^{-1}||\nabla_{\gammaup}\Theta_{\gammaup}||^{2}
\end{equation}
where $I$ is a real function. In case $U$ is a positive real
function $I=0.$
\end{lemma}

The spinor $\Theta_{\chi^{-}}$ satisfies
$D_{\chi^{-}}\Theta_{\chi^{-}}=0$ by the conformal transformation
formula. For it Eq.~\eqref{lich2} becomes after dividing out by
$c^{4}/16,$
\begin{equation} \label{lich1minus}
  2\Delta _{\chi^{-} }||r_{\text{in}}^{-2}\Theta_{\chi}||^{2}=R_{\chi^{-}}||r_{\text{in}}^{-2}\Theta_{\chi}||^{2}+4||\nabla
_{\chi^{-} }(r_{\text{in}}^{-2}\Theta_{\chi})||^{2}
\end{equation}

Using Eqs.~(\ref{done}-\ref{3laplnr}) we obtain
\scriptsize
\begin{multline}\label{aux1}
2\Delta _{\chi^{-}
}||r_{\text{in}}^{-2}\Theta_{\chi}||^{2}-4||\nabla _{\chi^{-}
}(r_{\text{in}}^{-2}\Theta_{\chi})||^{2}=-4r_{\text{in}}^{-4}||\Theta_{\chi}||^{2}\left\langle
\overline{\nabla \strut }\ln U^{-},
\overline{\nabla \strut }\ln
r_{\text{in}}\right\rangle _{\chi^{-}
}\\+2r_{\text{in}}^{-4}\Delta _{\chi^{-}
}||\Theta_{\chi}||^{2}-4(c^{4}/16)r_{\text{in}}^{-8} ||\nabla
_{\chi}
\Theta_{\chi}||^{2}-4r_{\text{in}}^{-4}\left\langle
\nabla
\ln r_{\text{in}},\nabla
||\Theta_{\chi}||^{2}\right\rangle_{\chi^{-}}
\end{multline}
\normalsize \,\, So Eq.~\eqref{lich1minus} gives
\scriptsize
\begin{equation}\label{onsigmaminus}
  2\Delta _{\chi^{-}
}||\Theta_{\chi}||^{2}=R_{\chi^{-}}||\Theta_{\chi}||^{2}+
4(M^{4}/16)r_{\text{in}}^{-4}||\nabla _{\chi}
\Theta_{\chi}||^{2}+4||\Theta_{\chi}||^{2}\left\langle
\overline{\nabla \strut }\ln U,
\overline{\nabla \strut }\ln
r_{\text{in}}\right\rangle _{\chi^{-} }+4\left\langle
\nabla
\ln r_{\text{in}},\nabla
||\Theta_{\chi}||^{2}\right\rangle_{\chi^{-}}
\end{equation}
\normalsize
Finally Eq.~\eqref{scalarchiminus} gives
\scriptsize
\begin{multline}\label{lich4}
  2\Delta _{\chi^{-}
}||\Theta_{\chi}||^{2}= 4(c^{4}/16)r_{\text{in}}^{-4}||\nabla
_{\chi}
\Theta_{\chi}||^{2}+4||\Theta_{\chi}||^{2}\left\langle
\overline{\nabla \strut }\ln U,
\overline{\nabla \strut }\ln
r_{\text{in}}\right\rangle _{\chi^{-} }+4\left\langle
\nabla
\ln r_{\text{in}},\nabla
||\Theta_{\chi}||^{2}\right\rangle_{\chi^{-}} \\
  +\sigma^{-2}(\zeta^{-})^{-1}\left(\mathcal{P}-Q_{\overline{g}}-U^{-1}\overline{\Delta }U
+\dfrac{1}{2}|\overline{
\nabla \strut }\ln U|^{2}-\left\langle
\overline{\nabla \strut }\ln U,\overline{\nabla \strut }\ln
f^{-}\right\rangle\right)||\Theta_{\chi}||^{2}
\end{multline}
\normalsize
Identity for $\Sigma^{-}$ is given in the following lemma.
\begin{lemma}
\scriptsize
\begin{multline}\label{idminus}
  \overline{\nabla \strut }_{\overline{\chi^{-}
}}\left(\sin^{2}\theta\left( 2\overline{
\nabla \strut }||\Theta _{\chi}||^{2}-2||\Theta _{\chi}||^{2}\overline{
\nabla \strut }\ln r_{\text{in}}+2|U|^{-3/4}\overline{\nabla \strut }\ln
r_{\text{in}}+||\Theta _{\chi}||^{2}\overline{\nabla
\strut }\ln U\right)-(||\Theta _{\chi
}||^{2}-1)\overline{\nabla\strut}\sin^{2}\theta\right)\\
 =\left(4(c^{4}/16)r_{\text{in}}^{-4}\sigma^{2}\zeta^{-} (\mathcal{R}+iI)-(3/2)L^{-}
\left\langle \overline{\nabla \strut }\ln U,\overline{
\nabla \strut }\ln f^{-}\right\rangle+i(3/4)|U|^{-3/4}
\left\langle \overline{\nabla \strut }\omega,\overline{
\nabla \strut }\ln f^{-}\right\rangle\right)\sigma^{-2}(\zeta^{-})^{-1}\sin^{2}\theta\\+
\sigma^{-2}(\zeta^{-})^{-1}\left(\mathcal{P}-Q_{\overline{g}}-(1/2)\left\langle
\overline{\nabla \strut }\ln U,\overline{\nabla \strut }\ln U\right\rangle-
(1/4)(4-3L^{-})\left\langle
\overline{\nabla \strut }\ln U,\overline{\nabla \strut }\ln
f^{-}\right\rangle\right)||\Theta_{\chi}||^{2}\sin^{2}\theta\\-2||\Theta
_{\chi}||^{2}\sigma^{-2}(\zeta^{-})^{-1}\overline{g}^{\theta\theta}\cos(2\theta)+\overline{\Delta
}_{\chi^{-}}\sin^{2}\theta
\end{multline}
\normalsize
where $I$ is as in \lemref{newlemma100} and $\mathcal{R}=
|U|^{-1}||\nabla_{\gammaup}\Theta _{\gammaup }||^{2}\geq 0.$
\end{lemma}
\begin{proof} Writing the 3-Laplacian $\Delta _{\chi^{-} }$ relative to
$\chi^{-} =\sigma^{2}\zeta^{-}
\overline{g}+Uf^{-}d\phi ^{2}$ in terms of the Laplacian of the 2-metric
$\overline{\chi^{-}}=\sigma^{2}\zeta^{-}\overline{g}$ we get \\
\scriptsize
$2\Delta_{\chi^{-}}||\Theta_{\chi} ||^{2}=2\overline{\Delta
}_{\overline{\chi^{-}}}||\Theta_{\chi} ||^{2}+\left\langle
\overline{\nabla \strut }\ln \left( Uf^{-}\right) ,
\overline{\nabla \strut }||\Theta_{\chi}  ||^{2}\right\rangle _{\chi^{-} }.$ We also have
$U^{-1}||\Theta _{\chi}||^{2}\overline{\Delta }_{\overline{
\chi^{-} }}U=\overline{\nabla \strut }_{\overline{\chi^{-} }}\left( ||\Theta _{\chi}||^{2}
U^{-1}\overline{\nabla \strut }U\right) -\left\langle
\overline{
\nabla \strut }||\Theta _{\chi}||^{2},\overline{\nabla \strut }\ln
U\right\rangle _{\chi^{-} }+\left\langle
\overline{\nabla
\strut }\ln U,\overline{\nabla \strut }\ln U\right\rangle
_{\chi^{-} }||\Theta _{\chi}||^{2}.$ So Eq.~\eqref{lich4} gives
\\
\begin{multline*}
  2\overline{\Delta }_{\overline{\chi^{-} }}||\Theta
_{\chi}||^{2}+\left\langle
\overline{\nabla \strut }\ln f^{-},\overline{\nabla \strut }
||\Theta _{\chi}||^{2}\right\rangle _{\chi^{-}}+\overline{\nabla
\strut }_{\overline{\chi^{-} }}\left( ||\Theta
_{\chi}||^{2}U^{-1}\overline{\nabla \strut }U\right)=
4(c^{4}/16)r_{\text{in}}^{-4}||\nabla _{\chi}
\Theta_{\chi}||^{2}+4||\Theta_{\chi}||^{2}\left\langle
\overline{\nabla \strut }\ln U,
\overline{\nabla \strut }\ln
r_{\text{in}}\right\rangle _{\chi^{-} }\\+4\left\langle
\nabla
\ln r_{\text{in}},\nabla
||\Theta_{\chi}||^{2}\right\rangle_{\chi^{-}}
  +\sigma^{-2}(\zeta^{-})^{-1}\left(\mathcal{P}-Q_{\overline{g}}-(1/2)\left\langle
\overline{\nabla \strut }\ln U,\overline{\nabla \strut }\ln U\right\rangle-\left\langle
\overline{\nabla \strut }\ln U,\overline{\nabla \strut }\ln
f^{-}\right\rangle\right)||\Theta_{\chi}||^{2} \Rightarrow
\end{multline*}
\begin{multline*} 2\overline{\Delta }_{\overline{\chi^{-}
}}||\Theta _{\chi}||^{2}-2\left\langle
\overline{\nabla \strut }\ln r_{\text{in}},\overline{\nabla \strut }
||\Theta _{\chi}||^{2}\right\rangle _{\chi^{-}}+\overline{\nabla
\strut }_{\overline{\chi^{-} }}\left( ||\Theta
_{\chi}||^{2}U^{-1}\overline{\nabla \strut }U\right)=
4(c^{4}/16)r_{\text{in}}^{-4}||\nabla _{\chi}
\Theta_{\chi}||^{2}+4||\Theta_{\chi}||^{2}\left\langle
\overline{\nabla \strut }\ln U,
\overline{\nabla \strut }\ln
r_{\text{in}}\right\rangle _{\chi^{-} }\\-\left\langle
\overline{\nabla \strut }\ln \sin^{2}\theta,\overline{\nabla \strut }
||\Theta _{\chi}||^{2}\right\rangle _{\overline{\chi^{-}}}
  +\sigma^{-2}(\zeta^{-})^{-1}\left(\mathcal{P}-Q_{\overline{g}}-(1/2)\left\langle
\overline{\nabla \strut }\ln U,\overline{\nabla \strut }\ln U\right\rangle-\left\langle
\overline{\nabla \strut }\ln U,\overline{\nabla \strut }\ln
f^{-}\right\rangle\right)||\Theta_{\chi}||^{2} \Rightarrow
\end{multline*}
\begin{multline*}
 2\overline{\Delta }_{\overline{\chi^{-} }}||\Theta
_{\chi}||^{2}+\overline{
\nabla \strut }_{\overline{\chi^{-} }}\left( -2||\Theta _{\chi}||^{2}\overline{
\nabla \strut }\ln r_{\text{in}}+2|U|^{-\dfrac{3}{4}}\overline{\nabla \strut }\ln
r_{\text{in}}+||\Theta _{\chi}||^{2}\overline{\nabla \strut }\ln
U\right) =\\ 4(c^{4}/16)r_{\text{in}}^{-4}||\nabla _{\chi}
\Theta_{\chi}||^{2}-(3/2)|U|^{-3/4}
\left\langle \overline{\nabla \strut }\ln |U|,\overline{
\nabla \strut }\ln r_{\text{in}}\right\rangle _{\chi^{-} }+4||\Theta_{\chi}||^{2}\left\langle
\overline{\nabla \strut }\ln U,
\overline{\nabla \strut }\ln
r_{\text{in}}\right\rangle _{\chi^{-} }\\-\left\langle
\overline{\nabla \strut }\ln \sin^{2}\theta,\overline{\nabla \strut }
||\Theta _{\chi}||^{2}\right\rangle _{\chi^{-}}
  +\sigma^{-2}(\zeta^{-})^{-1}\left(\mathcal{P}-Q_{\overline{g}}-(1/2)\left\langle
\overline{\nabla \strut }\ln U,\overline{\nabla \strut }\ln U\right\rangle-\left\langle
\overline{\nabla \strut }\ln U,\overline{\nabla \strut }\ln
f^{-}\right\rangle\right)||\Theta_{\chi}||^{2}
\end{multline*}
\normalsize
Multiplying both sides by $\sin^{2}\theta$ and using that $U,
r_{\text{in}}$ are functions of $r$ only we get
\scriptsize
\begin{multline*}
\overline{\nabla \strut }_{\overline{\chi^{-}
}}\left(\sin^{2}\theta\left( 2\overline{
\nabla \strut }||\Theta _{\chi}||^{2}-2||\Theta _{\chi}||^{2}\overline{
\nabla \strut }\ln r_{\text{in}}+2|U|^{-3/4}\overline{\nabla \strut }\ln
r_{\text{in}}+||\Theta _{\chi}||^{2}\overline{\nabla
\strut }\ln U\right)\right)-2\left\langle\overline{\nabla \strut
}\sin^{2}\theta,\overline{
\nabla \strut }||\Theta _{\chi}||^{2}\right\rangle_{\chi^{-}
}\\ =4\sin^{2}\theta (c^{4}/16)r_{\text{in}}^{-4}||\nabla _{\chi}
\Theta_{\chi}||^{2}-(3/2)\sin^{2}\theta|U|^{-3/4}
\left\langle \overline{\nabla \strut }\ln |U|,\overline{
\nabla \strut }\ln r_{\text{in}}\right\rangle _{\chi^{-} }+4\sin^{2}\theta||\Theta_{\chi}||^{2}\left\langle
\overline{\nabla \strut }\ln U,
\overline{\nabla \strut }\ln
r_{\text{in}}\right\rangle _{\chi^{-} }\\-\left\langle
\overline{\nabla \strut }\sin^{2}\theta,\overline{\nabla \strut }
||\Theta _{\chi}||^{2}\right\rangle _{\chi^{-}}
  +\sigma^{-2}(\zeta^{-})^{-1}\left(\mathcal{P}-Q_{\overline{g}}-(1/2)\left\langle
\overline{\nabla \strut }\ln U,\overline{\nabla \strut }\ln U\right\rangle-\left\langle
\overline{\nabla \strut }\ln U,\overline{\nabla \strut }\ln
f^{-}\right\rangle\right)||\Theta_{\chi}||^{2}\sin^{2}\theta
\Rightarrow
\end{multline*}
\begin{multline*}
 \overline{\nabla \strut }_{\overline{\chi^{-}
}}\left(\sin^{2}\theta\left( 2\overline{
\nabla \strut }||\Theta _{\chi}||^{2}-2||\Theta _{\chi}||^{2}\overline{
\nabla \strut }\ln r_{\text{in}}+2|U|^{-3/4}\overline{\nabla \strut }\ln
r_{\text{in}}+||\Theta _{\chi}||^{2}\overline{\nabla
\strut }\ln U\right)\right)-\overline{\nabla \strut }_{\overline{\chi^{-}
}}\left(||\Theta_{\chi}||^{2}\nabla \sin^{2}\theta\right)\\
 =4\sin^{2}\theta (c^{4}/16)r_{\text{in}}^{-4}||\nabla _{\chi}
\Theta_{\chi}||^{2}  -(3/2)\sin^{2}\theta|U|^{-3/4}
\left\langle \overline{\nabla \strut }\ln |U|,\overline{
\nabla \strut }\ln r_{\text{in}}\right\rangle _{\chi^{-} }+4\sin^{2}\theta||\Theta_{\chi}||^{2}\left\langle
\overline{\nabla \strut }\ln U,
\overline{\nabla \strut }\ln
r_{\text{in}}\right\rangle _{\chi^{-} }+\\
\sigma^{-2}(\zeta^{-})^{-1}\left(\mathcal{P}-Q_{\overline{g}}-(1/2)\left\langle
\overline{\nabla \strut }\ln U,\overline{\nabla \strut }\ln U\right\rangle-\left\langle
\overline{\nabla \strut }\ln U,\overline{\nabla \strut }\ln
f^{-}\right\rangle\right)||\Theta_{\chi}||^{2}\sin^{2}\theta-||\Theta
_{\chi}||^{2}\overline{\Delta }_{\chi^{-}}\sin^{2}\theta
\Rightarrow
\end{multline*}
\begin{multline*}
 \overline{\nabla \strut }_{\overline{\chi^{-}
}}\left(\sin^{2}\theta\left( 2\overline{
\nabla \strut }||\Theta _{\chi}||^{2}-2||\Theta _{\chi}||^{2}\overline{
\nabla \strut }\ln r_{\text{in}}+2|U|^{-3/4}\overline{\nabla \strut }\ln
r_{\text{in}}+||\Theta _{\chi}||^{2}\overline{\nabla
\strut }\ln U\right)-(||\Theta _{\chi
}||^{2}-1)\overline{\nabla\strut}\sin^{2}\theta\right)\\
 =\sigma^{-2}(\zeta^{-})^{-1}\left(4 (c^{4}/16)r_{\text{in}}^{-4}\sigma^{2}\zeta^{-}||\nabla _{\chi}
\Theta_{\chi}||^{2}  -(3/4)|U|^{-3/4}
\left\langle \overline{\nabla \strut }\ln U,\overline{
\nabla \strut }\ln f^{-}\right\rangle+i(3/4)|U|^{-3/4}
\left\langle \overline{\nabla \strut }\omega,\overline{
\nabla \strut }\ln f^{-}\right\rangle+\right.\\\left.4||\Theta_{\chi}||^{2}\left\langle
\overline{\nabla \strut }\ln U,
\overline{\nabla \strut }\ln
r_{\text{in}}\right\rangle +
\left(\mathcal{P}-Q_{\overline{g}}-(1/2)\left\langle
\overline{\nabla \strut }\ln U,\overline{\nabla \strut }\ln U\right\rangle-\left\langle
\overline{\nabla \strut }\ln U,\overline{\nabla \strut }\ln
f^{-}\right\rangle\right)||\Theta_{\chi}||^{2}\right)\sin^{2}\theta-(||\Theta
_{\chi}||^{2}-1)\overline{\Delta }_{\chi^{-}}\sin^{2}\theta
\end{multline*}
\normalsize
Using $|U|^{-3/4}=L^{-}||\Theta_{\chi}||^{2}$ and
Eq.~\eqref{lapsin2} we get
\scriptsize
\begin{multline*}
 \overline{\nabla \strut }_{\overline{\chi^{-}
}}\left(\sin^{2}\theta\left( 2\overline{
\nabla \strut }||\Theta _{\chi}||^{2}-2||\Theta _{\chi}||^{2}\overline{
\nabla \strut }\ln r_{\text{in}}+2|U|^{-3/4}\overline{\nabla \strut }\ln
r_{\text{in}}+||\Theta _{\chi}||^{2}\overline{\nabla
\strut }\ln U\right)-(||\Theta _{\chi
}||^{2}-1)\overline{\nabla\strut}\sin^{2}\theta\right)\\
 =\sigma^{-2}(\zeta^{-})^{-1}\left(4 (c^{4}/16)r_{\text{in}}^{-4}\sigma^{2}\zeta^{-}||\nabla _{\chi}
\Theta_{\chi}||^{2}  -(3/2)L^{-}||\Theta_{\chi}||^{2}
\left\langle \overline{\nabla \strut }\ln U,\overline{
\nabla \strut }\ln f^{-}\right\rangle+i(3/4)|U|^{-3/4}
\left\langle \overline{\nabla \strut }\omega,\overline{
\nabla \strut }\ln f^{-}\right\rangle+\right.\\\left.4||\Theta_{\chi}||^{2}\left\langle
\overline{\nabla \strut }\ln U,
\overline{\nabla \strut }\ln
r_{\text{in}}\right\rangle +
\left(\mathcal{P}-Q_{\overline{g}}-(1/2)\left\langle
\overline{\nabla \strut }\ln U,\overline{\nabla \strut }\ln U\right\rangle-
(1/4)(4-3L^{-})\left\langle
\overline{\nabla \strut }\ln U,\overline{\nabla \strut }\ln
f^{-}\right\rangle\right)||\Theta_{\chi}||^{2}\right)\sin^{2}\theta\\-2||\Theta
_{\chi}||^{2}\sigma^{-2}(\zeta^{-})^{-1}\overline{g}^{\theta\theta}\cos(2\theta)+\overline{\Delta
}_{\chi^{-}}\sin^{2}\theta
\end{multline*}
\normalsize
Since \scriptsize $4\sigma^{-2}(\zeta^{-})^{-1}\left\langle
\overline{\nabla \strut }\ln U,
\overline{\nabla \strut }\ln
r_{\text{in}}\right\rangle=-2\sigma^{-2}(\zeta^{-})^{-1}\left\langle
\overline{\nabla \strut }\ln U,
\overline{\nabla \strut }\ln f\right\rangle=-2(c^{4}/16)r_{\text{in}}^{-4}\left\langle
\overline{\nabla \strut }\ln U,
\overline{\nabla \strut }\ln
r_{\text{in}}\right\rangle _{\overline{\chi }}$ \normalsize we get
Eq.~\eqref{idminus} using \lemref{newlemma100}.
\end{proof}

\section{Main theorem}
We separate the positive part of the contribution of the integral
of \scriptsize $||\Theta
_{\chi}||^{2}\sigma^{-2}(\zeta^{\pm})^{-1}\overline{g}^{\theta\theta}\cos(2\theta)$
\normalsize on $\Sigma^{\pm}.$ Although this integral will be zero in case
$||\Theta_{\chi}||^{2}$ is constant the integrand is not
identically zero even in the case of Kerr-Newman solution. This
follows from Eq.~\eqref{lapsin2}. Let
\begin{equation} \label{Deltasinave}
\mathcal{A}\left(r_{0}\right)=\dfrac{\int\limits_{r =r_{0}}
\sigma^{3/2}||\Theta_{\vartheta} ||^{2}\cos(2\theta)d\theta
}{\int\limits_{r=r_{0}}\sigma^{3/2}||\Theta_{\vartheta}||^{2}\sin^{2}\theta
d\theta }
\end{equation}
As usual we are interested only for $r_{0}>M+c.$ We recall that
$\sigma^{3/2}||\Theta_{\vartheta}
||^{2}=X_{\text{K}}^{1/8}\zeta^{-1/4}X^{5/8}f^{-3/4}||\Theta_{\widehat{g}}||^{2}
=X_{\text{K}}^{1/8}O\left((\sin\theta)^{-1/4}\right)$ which is
integrable. We show
\begin{equation}\label{mathcalA}
  \int\limits_{\left( r_{1},r_{2}\right)
,\overline{\chi^{\pm}}}||\Theta
_{\chi}||^{2}\sigma^{-2}(\zeta^{\pm})^{-1}\overline{g}^{\theta\theta}\cos(2\theta)=\int\limits_{\left(
r_{1},r_{2}\right)
,\overline{\chi^{\pm}}}\sin^{2}\theta||\Theta_{\chi}||^{2}\sigma^{-2}(\zeta^{\pm})^{-1}\overline{g}^{\theta\theta}\mathcal{A}
\end{equation}
LHS$ =\iint\limits_{\left(
r_{1},r_{2}\right)}|U|^{-3/4}||\Theta_{\vartheta}||^{2}
\sigma^{3/2-2}(\zeta^{\pm})^{-1}\overline{g}^{\theta\theta}\cos(2\theta)
\sqrt{\chi^{\pm}_{rr}\chi^{\pm}_{\theta\theta }}drd\theta \\
=\iint\limits_{\left(
r_{1},r_{2}\right)}|U|^{-3/4}\sigma^{3/2}||\Theta_{\vartheta}||^{2}\cos(2\theta)
(r^{2}-2M^{\prime}r+a^{2})^{-1/2} drd\theta\\
=\int_{r_{1}}^{r_{2}}|U|^{-3/4}(r^{2}-2M^{\prime}r+a^{2})^{-1/2}\left(\int\sigma^{3/2}||\Theta_{\vartheta}||^{2}\cos(2\theta)
d\theta\right)dr \\
=\int_{r_{1}}^{r_{2}}|U|^{-3/4}(r^{2}-2M^{\prime}r+a^{2})^{-1/2}\mathcal{A}\left(\int\sigma^{3/2}||\Theta_{\vartheta}||^{2}\sin^{2}\theta
d\theta\right)dr\\ =\iint\limits_{\left(
r_{1},r_{2}\right)}\sin^{2}\theta|U|^{-3/4}\sigma^{3/2}||\Theta_{\vartheta}||^{2}(r^{2}-2M^{\prime}r+a^{2})^{-1/2}\mathcal{A}drd\theta
\\
=\iint\limits_{\left(
r_{1},r_{2}\right)}\sin^{2}\theta|U|^{-3/4}\sigma^{3/2}||\Theta_{\vartheta}||^{2}(r^{2}-2M^{\prime}r+a^{2})^{-1/2}\mathcal{A}\sqrt{\chi^{rr}\chi^{\theta\theta
}}\sqrt{\chi _{rr}\chi _{\theta\theta }}drd\theta
\\
 =\int\limits_{\left(
r_{1},r_{2}\right)
,\overline{\chi^{\pm}}}\sin^{2}\theta||\Theta_{\chi}||^{2}\sigma^{-2}(\zeta^{\pm})^{-1}\overline{g}^{\theta\theta}\mathcal{A}
.$ Having checked Eq.~\eqref{mathcalA} we now decompose
$\mathcal{A}$ into positive and negative parts. Positive part we
add with $Q_{\overline{g}}$ since there is a negative sign in
front of $2||\Theta
_{\chi}||^{2}(\zeta^{-})^{-1}\overline{g}^{\theta\theta}\cos(2\theta)$
in Eqs.~(\ref{id},\ref{idminus}). We define
\begin{equation}\label{widehatQ}
\widehat{Q}_{\overline{g}}=Q_{\overline{g}}+2\overline{g}^{\theta\theta}\mathcal{A}^{+}
\end{equation}
$\widehat{Q}_{\overline{g}}$ is nonnegative. This will be
important in the proof of the main theorem. We define $Q_{\Pi
}=\left(4\left\langle \overline{\nabla \strut }
\ln \left( X_{\text{K}}/\sqrt{\rho }\right) ,\overline{\nabla \strut }\ln
\sigma \right\rangle_{\Pi}
+\overline{g}_{\theta\theta}f_{\text{em}}\right)^{-}.$ As explained before \lemref{lemmarchi}, $Q_{\Pi
}$ is well-defined for $r>M+c.$
Then from Eqs.~(\ref{ginrtheta},\ref{gthetatheta},\ref{Qg})
we get $Q_{\overline{g}} =\overline{g}^{\theta\theta}Q_{\Pi }.$
Similarly we write Eq.~\eqref{widehatQ} as
\begin{equation}\label{widehatQpi}
\widehat{Q}_{\overline{g}}=\overline{g}^{\theta\theta}\widehat{Q}_{\Pi }
\end{equation}
where $\widehat{Q}_{\Pi}=Q_{\Pi}+2\mathcal{A}^{+}.$ We now define
an average of $\widehat{Q}_{\Pi}$ on a $r=\text{constant}$ loop on
a $\phi=\text{constant}$ surface.
\begin{equation} \label{Qpiave}
Q_{\text{ave}}(r_{0})=\dfrac{\int\limits_{r =r_{0}}\sigma^{3/2}
\widehat{Q}_{\Pi }||\Theta_{\vartheta} ||^{2}\sin^{2}\theta
d\theta}{\int\limits_{r=r_{0}}\sigma^{3/2}||\Theta_{\vartheta}||^{2}\sin^{2}\theta
d\theta}
\end{equation}
Thus $Q_{\text{ave}}(r_{0})=\dfrac{\int\limits_{r
=r_{0}}\sigma^{3/2} Q_{\Pi }||\Theta_{\vartheta}
||^{2}\sin^{2}\theta
d\theta}{\int\limits_{r=r_{0}}\sigma^{3/2}||\Theta_{\vartheta}||^{2}\sin^{2}\theta
d\theta}+2\mathcal{A}^{+}.$ Since $\int\limits_{r
=r_{0}}\sigma^{3/2}||\Theta_{\vartheta}||^{2}\cos(2\theta)d\theta=\int\limits_{r
=r_{0}}\left(1+O(r^{-1})\right)\cos(2\theta)d\theta=O(r^{-1})$ as
$r\rightarrow
\infty,$ we have
\begin{equation}\label{cos2theta}
Q_{\text{ave}}(r)=O(r^{-1})
\end{equation}
Also as $\sin\theta \rightarrow 0$ and $r\downarrow M+c,$
$Q_{\text{ave}}(r)$ is bounded because $Q_{\Pi}$ is at worst
$O\left((\sin\theta)^{-2}\right)$ and
$\sigma^{3/2}||\Theta_{\vartheta}
||^{2}=X_{\text{K}}^{1/8}O\left((\sin\theta)^{-1/4}\right).$ In fact we
have
\begin{equation*}
  4\left\langle \overline{\nabla \strut }\ln \left(
X_{\text{K}}/\sqrt{\rho}\right),\overline{\nabla \strut }\ln \sigma \right\rangle =\left\{
\begin{array}{c}
\overline{g}^{\theta\theta}O( r^{-2}) \ \text{as }r \rightarrow \infty , \\
\overline{g}^{\theta\theta}O(\sin\theta) \ \text{as }\theta \rightarrow 0 \text{ or }\pi, \\
\overline{g}^{\theta\theta}O( 1)
\text{ }\ \text{as }r \rightarrow M+c,\text{ }\theta \neq 0,\pi
\end{array}
\right.
\end{equation*}
where, we recall from Eq.~\eqref{gthetatheta}, that $\overline{g}^{\theta\theta}=\Omega^{-1}(r^{2}-2M^{\prime}r
+(M^{2}-\mathfrak{e}^{2}-\mathfrak{m}^{2})\sin^{2}\theta+a^{2}\cos^{2}\theta)^{-1}.$ $\overline{g}^{\theta\theta}$ is well defined for
$r>M+c.$ This factor does not occur in the inner product relative to the metric $\Pi$ and hence in $Q_{\Pi }$ and $Q_{\text{ave}}.$
Also $Q_{\text{ave}}(r)$ is not defined at isolated points where
$\int\limits_{r=\hat{r}}\sigma^{3/2}||\Theta_{\vartheta}
||^{2}\sin^{2}\theta d\theta=0$ but these are removable and $Q_{\text{ave}}(r)$ remains continuous after redefinition.
 Such a value exists only if the known spinor $\Theta_{\widehat{g}}$ vanishes on the entire $r=\hat{r}$ loop.

\begin{lemma} \label{lemma11.2}
Suppose $U=U\left(r\right)$ and
\scriptsize
\begin{equation} \label{qdplus}
\left(r^{2}-2M^{\prime}r+a^{2}\right) \left( \dfrac{d\ln U}{dr
}\right)
^{2}+(4-3L_{\text{ave}})\sqrt{r^{2}-2M^{\prime}r+a^{2}}\dfrac{d\ln
U}{dr }+2Q_{\text{ave}}=0
\end{equation}
\begin{equation} \label{parentode}
\text{\normalsize Then } \scriptsize \iint\limits_{\Sigma^{+}_{2},\overline{\chi}}\left(2\left\langle\nabla \ln U,\nabla
\ln U\right\rangle+(4-3L)\left\langle
\overline{\nabla \strut }\ln U,\overline{\nabla \strut }\ln \ f\right\rangle+4 \widehat{Q}_{\overline{g}}\right)
\sigma^{-2}\zeta^{-1}\sin^{2}\theta||\Theta_{\chi}||^{2}=0
\end{equation}
\end{lemma}
\begin{proof} Using Eqs.~(\ref{spchi},\ref{chi},\ref{eta}) we see that
the LHS of Eq.~\eqref{parentode} is \\
\scriptsize
$\iint\limits_{\left( r_{1},r_{2}\right) }\sin^{2}\theta\left(2
\left(r^{2}-2M^{\prime}r+a^{2}\right) \left( \dfrac{d\ln U}{dr }\right)
^{2}+(4-3L)\left(r^{2}-2M^{\prime}r+a^{2}\right) \dfrac{d\ln U}{dr
}\dfrac{\partial\ln f }{\partial r}+4 \widehat{Q}_{\Pi }\right)
\overline{g}^{\theta\theta}
\sigma^{-2}\zeta ^{-1}||\Theta _{\chi }||^{2}
\sqrt{\chi _{rr}\chi _{\theta\theta }}drd\theta
$ \\ $=\iint\limits_{\left( r_{1},r_{2}\right)
}\sin^{2}\theta\left(2
\left(r^{2}-2M^{\prime}r+a^{2}\right) \left( \dfrac{d\ln U}{dr }\right)
^{2}+(4-3L)\left(r^{2}-2M^{\prime}r+a^{2}\right) \dfrac{d\ln U}{dr
}\dfrac{\partial\ln f }{\partial r}+4 \widehat{Q}_{\Pi }\right)
|U|^{-3/4}\sigma^{3/2}||\Theta_{\vartheta} ||^{2}\\
(r^{2}-2M^{\prime}r+a^{2})^{-1/2}drd\theta .$
\normalsize \,\, We use the definition of $Q_{\text{ave}}$ given in
Eq.~\eqref{Qpiave}. Also from
\eqref{lave} and noting that $(\partial\ln f/\partial r )=2(r^{2}-2M^{\prime}r+a^{2})^{-1/2}$ we get
\scriptsize
\begin{equation*}
\int L\sigma^{3/2}\left(r^{2}-2M^{\prime}r+a^{2}\right) \dfrac{d\ln U}{dr
}\dfrac{\partial\ln f }{\partial
r}||\Theta_{\vartheta}||^{2}\sin^{2}\theta d\theta =
2L_{\text{ave}}\sqrt{r^{2}-2M^{\prime}r+a^{2}}
\dfrac{d\ln U}{dr }\int \sigma^{3/2}
||\Theta_{\vartheta}||^{2}\sin^{2}\theta d\theta
\end{equation*}
\normalsize
So replacing $\widehat{Q}_{\Pi }$ and $L$ by their respective
average values $Q_{\text{ave}}$ and $L_{\text{ave}}$ we write the
LHS of Eq.~\eqref{parentode} as \\
\scriptsize
 $\iint\limits_{\left(
r_{1},r_{2}\right) }\sin^{2}\theta\left(2
\left(r^{2}-2M^{\prime}r+a^{2}\right) \left( \dfrac{d\ln U}{dr }\right)
^{2}+2(4-3L_{\text{ave}})\sqrt{r^{2}-2M^{\prime}r+a^{2}}
\dfrac{d\ln U}{dr }+4 Q_{\text{ave}}\right) |U|^{-3/4}\sigma^{3/2}||\Theta
||^{2}\\ (r^{2}-2M^{\prime}r+a^{2})^{-1/2}drd\theta$
\normalsize which is $0$
by Eq.~\eqref{qdplus}.
\end{proof}

Thus if $U$ is a solution of \eqref{qdplus} then integrating
\eqref{id} we get
\scriptsize
\begin{equation*}
\iint\limits_{\Sigma^{+}_{2},\overline{\chi}}\left(\sin^{2}\theta\left(4||\nabla _{\chi }\Theta _{\chi }||^{2}
+||\Theta_{\chi}||^{2}\sigma^{-2}\zeta^{-1}\left(\mathcal{P}+2\mathcal{A}^{-}\overline{g}^{\theta\theta}-(3/4)iL\left\langle\overline{\nabla
\strut }\omega ,\overline{\nabla \strut }\ln f\right\rangle\right)\right)\right)=
\lim\limits_{r_{1}\rightarrow
M+c}\int\limits_{r=r_{1}}\mathscr{B}d\theta
+\lim\limits_{r_{2}\rightarrow
\infty}\int\limits_{r=r_{2}}\mathscr{B}d\theta
\end{equation*}
\normalsize
where we wrote the boundary integrands with respect to outward
normals by $\mathscr{B}.$ Before computing $\mathscr{B}$ we cannot
give rigorous details. However we state roughly our plan. The part
of the axis given by $\theta=0$ or $\theta=\pi$ alone (alone means
in the region $r>M+c$ as explained after Eq.~\eqref{c}) does not
appear as a limiting boundary. If we consider a
$\theta=\text{constant}=\theta_{0}$ line for $\theta_{0}$ close to
$0$ or $\pi$ as a boundary then the LHS of Eq.~\eqref{id} gives on
this boundary
\scriptsize $\int\limits_{\theta_{0}}
\left\langle 2\sin^{2}\theta\overline{
\nabla \strut }||\Theta _{\chi }||^{2}-(||\Theta _{\chi }||^{2}-1)\overline{
\nabla \strut }\sin^{2}\theta,\overline{n}\right\rangle_{\overline{\chi}}$
\normalsize
where $\overline{n}$ is the unit normal form on this line. All
other terms are orthogonal to $\overline{n}.$ The integral
vanishes as $\theta_{0}\rightarrow 0 \text{ or } \pi$ because it
is bounded by $C\sin\theta
\int(r^{2}-2M^{\prime}r+a^{2})^{-1/2}r^{-1}dr$ for some constant
$C.$ $r^{-1}$ factor comes from $||\Theta _{\chi }||^{2}-1$ for
large values of $r.$ This factor is necessary only to make the
constant $C$ independent of $r.$ But we can also take the
$\theta_{0}\rightarrow 0
\text{ or } \pi$ limit for a fixed large $r$ first and then let
$r\rightarrow \infty.$ The boundary term from $\overline{\Delta
}_{\overline{\chi}}\sin^{2}\theta$ in the RHS of Eq.~\eqref{id}
vanishes.

Now we want to compute the boundary integrals. But in order to
evaluate the boundary integral at $r=M+c$ and not on the axis we
are forced to consider the identity in Eq.~\eqref{idminus} on
$\Sigma^{-}$ and to match $U$ suitably across the smooth parts of
the limiting surface at $r=M+c.$ We proceed as follows.

A solution of Eq.~\eqref{qdplus} for $4-3L_{\text{ave}}\geq 0$ is
given by the first case of the following equation.
\begin{equation}\label{cqdp}
\dfrac{d\ln
U}{dr}=
  \begin{cases}
\dfrac{-2+(3/2)L_{\text{ave}}+\sqrt{(2-(3/2)L_{\text{ave}})^{2}-2Q_{\text{ave}}}}{\sqrt{r^{2}-2M^{\prime}r+a^{2}}}&
\text{ if $4-3L_{\text{ave}}\geq 0,$} \\
i\sqrt{2Q_{\text{ave}}/(r^{2}-2M^{\prime}r+a^{2})}& \text{
otherwise }
\end{cases}
\end{equation}

For this solution for $r>M+c,$ $\re\dfrac{d\ln U}{dr }\leq 0.$
Writing for this solution
\scriptsize
\begin{equation}\label{extp}
-\iint\limits_{\Sigma^{+}_{2},\overline{\chi}}\sigma^{-2}\zeta^{-1}\sin^{2}\theta\left(
(1/2)\left\langle\nabla \ln U,\nabla
\ln U\right\rangle+(1/4)(4-3L)\left\langle
\overline{\nabla \strut }\ln U,\overline{\nabla \strut }\ln \ f\right\rangle+\widehat{Q}_{\overline{g}}
\right)||\Theta_{\chi}||^{2}=i\iint\limits_{\Sigma^{+}_{2},\overline{\chi}}
I_{1}\sigma^{-2}\zeta^{-1}\overline{g}^{\theta\theta}\sin^{2}\theta
||\Theta_{\chi}||^{2}
\end{equation}
\normalsize
we see that $I_{1}$ is a real function on $\Sigma^{+}.$ Absorbing
the other pure imaginary (or zero) term of Eq.~\eqref{id} we
define
\begin{equation}\label{Iout}
i\mathcal{I}_{\text{out}}\zeta^{-1}=iI_{1}\zeta^{-1}\overline{g}^{\theta\theta}-(3/4)\zeta^{-1}iL|U|^{-3/4}\left\langle\overline{\nabla
\strut }\omega ,\overline{\nabla \strut }\ln f\right\rangle
\end{equation}

Similarly on $\Sigma^{-}$ we seek $U$ such that
\begin{equation} \label{newm}
2\left\langle\nabla \ln U,\nabla
\ln U\right\rangle+(4-3L^{-})\left\langle
\overline{\nabla \strut }\ln U,\overline{\nabla \strut }\ln \ f^{-}\right\rangle+4 \widehat{Q}_{\overline{g}}=0
\end{equation}
for $4-3L_{\text{ave}}\geq 0.$

Since $(\partial \ln f^{-}/\partial
r)=-2(r^2-2M^{\prime}r+a^{2})^{-1/2}$ this time we need
\begin{equation} \label{qdm}
\left(r^{2}-2M^{\prime}r+a^{2}\right) \left( \dfrac{d\ln U}{dr
}\right) ^{2}-(4-3L^{-}_{\text{ave}})\dfrac{d\ln U}{dr
}+2Q_{\text{ave}}=0
\end{equation}

A solution of Eq.~\eqref{qdm} for $4-3L^{-}_{\text{ave}}\geq 0$ is
given by the first case of
\begin{equation}\label{cqdm}
\dfrac{d\ln
U}{dr}=
  \begin{cases}
\dfrac{2-(3/2)L^{-}_{\text{ave}}-\sqrt{(2-(3/2)L^{-}_{\text{ave}})^{2}-2Q_{\text{ave}}}}{\sqrt{r^{2}-2Mr}}&
\text{ if $4-3L^{-}_{\text{ave}}\geq 0,$} \\
-i\sqrt{2Q_{\text{ave}}/(r^{2}-2Mr)}& \text{ otherwise }
\end{cases}
\end{equation}

For this solution for $r>M+c,$ $\re\dfrac{d\ln U}{dr }\geq 0.$ In
particular $\re L^{-}
\left\langle \overline{\nabla \strut }\ln U,\overline{
\nabla \strut }\ln f^{-}\right\rangle\leq 0.$ Thus in Eq.~\eqref{idminus} for
$\Sigma^{-}$ we write
\scriptsize
\begin{multline}\label{extm}
\iint\limits_{\Sigma^{-}_{2},\overline{\chi^{-}}}\sigma^{-2}(\zeta^{-})^{-1}\sin^{2}\theta\left(4(c^{4}/16)r_{\text{in}}^{-4}\sigma^{2}\zeta^{-}
\left(\mathcal{R}+iI\right)
-(1/2)\left\langle\nabla \ln U,\nabla
\ln U\right\rangle-(1/4)(4-3L^{-})\left\langle
\overline{\nabla \strut }\ln U,\overline{\nabla \strut }\ln
f^{-}\right\rangle-\widehat{Q}_{\overline{g}}\right. \\\left.
-(3/2)L^{-}
\left\langle \overline{\nabla \strut }\ln U,\overline{
\nabla \strut }\ln f^{-}\right\rangle+i(3/4)L^{-}
\left\langle \overline{\nabla \strut }\omega,\overline{
\nabla \strut }\ln f^{-}\right\rangle
\right)||\Theta_{\chi}||^{2}=\iint\limits_{\Sigma^{-}_{2},\overline{\chi^{-}}}
\sigma^{-2}(\zeta^{-})^{-1}\sin^{2}\theta
\left(\mathcal{R}_{\text{in}}+i\mathcal{I}_{\text{in}}\right)||\Theta_{\chi}||^{2}
\end{multline}
\normalsize
where both $\mathcal{R}_{\text{in}}$ and $\mathcal{I}_{\text{in}}$
are real and $\mathcal{R}_{\text{in}}\geq 0.$

Since on $\Sigma^{-},$ $|U|^{-3/4}=L^{-}||\Theta_{\chi}||^{2}$ and
$||\Theta_{\chi}||^{2}=\sigma^{3/2}|U|^{-3/4}||\Theta_{\vartheta}||^{2}$
we have $\sigma^{3/2}L^{-}||\Theta_{\vartheta}||^{2}=1.$ Since
$||\Theta_{\vartheta}||^{2}$ is continuous across $r=M+c,$ at
$r=M+c,$ $L^{-}=L$ and hence $L^{-}_{\text{ave}}=L_{\text{ave}}.$
It now follows from Eqs.~(\ref{cqdp},\ref{cqdm}) that

\begin{equation}\label{match1}
  \left.\dfrac{d\ln U}{dr_{\text{out} }}\right\vert_{r=M+c}=\left.\dfrac{d\ln U}{dr_{\text{in} }}\right\vert_{r=M+c}
\end{equation}
The above equation holds even before we match $U$ on both
$\Sigma^{\pm}_{2}$ at $r=M+c$ using the constant of integration.

\begin{lemma} \label{lemma11.3}
Let $n_{\overline{\chi} }$ be the unit normal form relative to the
metric $\overline{\chi} $ on the loop $r =r_{0}$ and the
corresponding vector points in the direction of decreasing $r.$
Let $U=U(r).$ For $M+c<r_{0}<\infty ,$
\scriptsize
\begin{multline} \label{boundary}
\oint\limits_{r_{0},\overline{\chi} }\left\langle
2\sin^{2}\theta\overline{
\nabla \strut }||\Theta _{\chi }||^{2}+
2\sin^{2}\theta||\Theta _{\chi }||^{2}\overline{
\nabla \strut }\ln r_{\text{out}}
-2\sin^{2}\theta |U|^{-3/4}\overline{\nabla \strut }\ln
r_{\text{out}}+\sin^{2}\theta||\Theta _{\chi
}||^{2}\overline{\nabla
\strut }\ln U,n_{\overline{\chi}}
\right\rangle_{\overline{\chi}} =
\dfrac{1}{|U(r_{0})|^{3/4}}\\
\int\limits_{r=r_{0}}
\left(\sigma^{3/2}\left(||\Theta ||^{2}\left((1/2)r_{\text{out}}\dfrac{d\ln
U}{dr_{\text{out}}}-
 i(3/2)\sqrt{r^{2}-2M^{\prime}r+a^{2}}\dfrac{d
\omega}{dr}-2\right)-
2\sqrt{r^{2}-2M^{\prime}r+a^{2}}\dfrac{\partial ||\Theta
||^{2}}{\partial r}\right) +2\right)\sin^{2}\theta d\theta
\end{multline}
\normalsize
Similarly on $\Sigma^{-}$ with $n_{\overline{\chi^{-}} }$ being
the unit normal form relative to the metric $\overline{\chi^{-}} $
on the loop $r =r_{0}$ and the corresponding vector pointing in
the direction of decreasing $r,$
\scriptsize
\begin{multline} \label{innerboundary}
\oint\limits_{r_{0},\overline{\chi} }\left\langle
2\sin^{2}\theta\overline{
\nabla \strut }||\Theta _{\chi }||^{2}+
2\sin^{2}\theta||\Theta _{\chi }||^{2}\overline{
\nabla \strut }\ln r_{\text{in}}
-2\sin^{2}\theta |U|^{-3/4}\overline{\nabla \strut }\ln
r_{\text{in}}+\sin^{2}\theta||\Theta _{\chi
}||^{2}\overline{\nabla
\strut }\ln U,n_{\overline{\chi^{-}}}
\right\rangle_{\chi^{-}} =
-\dfrac{1}{|U(r_{0})|^{3/4}} \\
\int\limits_{r=r_{0}}
\left(\sigma^{3/2}\left(||\Theta ||^{2}\left((1/2)r_{\text{in}}\dfrac{d\ln U}{dr_{\text{in}}}+ i(3/2)\sqrt{r^{2}-2M^{\prime}r+a^{2}}\dfrac{d
\omega}{dr}-2\right)+
2\sqrt{r^{2}-2M^{\prime}r+a^{2}}\dfrac{\partial ||\Theta
||^{2}}{\partial r}\right) +2\right)\sin^{2}\theta d\theta
\end{multline}
\normalsize
In case $U$ is real near infinity, $U^{-1}$ is bounded and $(d\ln
U/dr)=o\left(r^{-1}\right) $ as $r
\rightarrow \infty,$ right hand sides of Eqs.~(\ref{boundary},\ref{innerboundary}) vanish as
$r_{0} \rightarrow \infty.$
\end{lemma}
\begin{proof} First we note that \scriptsize $\overline{
\nabla \strut }||\Theta _{\chi }||^{2}=\overline{
\nabla \strut }\left(\sigma^{3/2}|U|^{-3/4}||\Theta_{\vartheta}||^{2}\right)=\overline{
\nabla \strut
}\left(\sigma^{3/2}U^{-3/4}e^{i(3/4)\omega}||\Theta_{\vartheta}||^{2}\right)\\
=\sigma^{3/2}e^{i(3/4)\omega}\left((3/2)U^{-3/4}||\Theta_{\vartheta}||^{2}\overline{
\nabla \strut }\ln\sigma
-(3/4)U^{-3/4}||\Theta_{\vartheta}||^{2}\overline{
\nabla \strut }\ln U+i(3/4)U^{-3/4}||\Theta_{\vartheta}||^{2}\overline{
\nabla \strut }\omega + U^{-3/4}\overline{
\nabla \strut }||\Theta_{\vartheta}||^{2}\right)\\
=(3/2)\sigma^{3/2}|U|^{-3/4}||\Theta_{\vartheta}||^{2}\overline{
\nabla \strut }\ln\sigma
-(3/4)\sigma^{3/2}|U|^{-3/4}||\Theta_{\vartheta}||^{2}\overline{
\nabla \strut }\ln U+i(3/4)\sigma^{3/2}|U|^{-3/4}||\Theta_{\vartheta}||^{2}\overline{
\nabla \strut }\omega + \sigma^{3/2}|U|^{-3/4}\overline{
\nabla \strut }||\Theta_{\vartheta}||^{2}.$ \normalsize Now $n_{\overline{\chi} ,r }=-\sqrt{\chi
_{rr} }=-\sigma\sqrt{\zeta
\overline{g}_{rr }}.$  So for any differentiable
function $F$ on the $r=r_{0}$ loop,
\scriptsize $\left\langle\overline{\nabla
\strut}F,n_{\overline{\chi}}\right\rangle_{\overline{\chi}}\sqrt{\overline{\chi}_{\theta\theta}}
=\overline{\chi}^{rr}\dfrac{\partial F}{\partial r}
n_{\overline{\chi} ,r }\sqrt{\overline{\chi}_{\theta\theta}}=
-\sqrt{\overline{g}^{rr}}\dfrac{\partial F}{\partial r}
\sqrt{\overline{g}_{\theta\theta}}=
-\sqrt{r^{2}-2M^{\prime}r+a^{2}}\dfrac{\partial F}{\partial r}.$
\normalsize
Thus using $||\Theta _{\chi
}||^{2}=\sigma^{3/2}|U|^{-3/4}||\Theta_{\vartheta}||^{2}$ we get
\scriptsize
\begin{multline*}
\oint\limits_{r_{0},\overline{\chi} }\left\langle
2\sin^{2}\theta\overline{
\nabla \strut }||\Theta _{\chi }||^{2}+
2\sin^{2}\theta||\Theta _{\chi }||^{2}\overline{
\nabla \strut }\ln r_{\text{out}}
-2\sin^{2}\theta |U|^{-3/4}\overline{\nabla \strut }\ln
r_{\text{out}}+\sin^{2}\theta||\Theta _{\chi
}||^{2}\overline{\nabla
\strut }\ln U,n_{\overline{\chi}}
\right\rangle_{\overline{\chi}} = \\
-\dfrac{\sqrt{r_{0}^{2}-2M^{\prime}r_{0}+a^{2}}}{|U(r_{0})|^{3/4}}
\int\limits_{r=r_{0}}
\left(\sigma^{3/2}\left(||\Theta_{\vartheta} ||^{2}\left(-(1/2)\dfrac{d\ln U}{dr}+ i(3/2)\dfrac{d
\omega}{dr}+2\dfrac{d\ln r_{\text{out}} }{dr }\right)+
2\dfrac{\partial ||\Theta_{\vartheta} ||^{2}}{\partial r}\right)
-2\dfrac{d\ln r_{\text{out}} }{dr }\right)\sin^{2}\theta d\theta
=\\ \dfrac{1}{|U(r_{0})|^{3/4}}
\int\limits_{r=r_{0}}
\left(\sigma^{3/2}\left(||\Theta_{\vartheta} ||^{2}\left((1/2)r_{\text{out}}\dfrac{d\ln
U}{dr_{\text{out}}}-
 i(3/2)\sqrt{r^{2}-2M^{\prime}r+a^{2}}\dfrac{d
\omega}{dr}-2\right)-
2\sqrt{r^{2}-2M^{\prime}r+a^{2}}\dfrac{\partial
||\Theta_{\vartheta} ||^{2}}{\partial r}\right)
+2\right)\sin^{2}\theta d\theta
\end{multline*}
\normalsize
 Similarly on
$\Sigma^{-}$ with $n_{\overline{\chi^{-}} }$ being the unit normal
form relative to the metric $\overline{\chi^{-}} $ on the loop $r
=r_{0}$ and the corresponding vector pointing in the direction of
decreasing $r,$
\scriptsize
\begin{multline*}
\oint\limits_{r_{0},\overline{\chi} }\left\langle
2\sin^{2}\theta\overline{
\nabla \strut }||\Theta _{\chi }||^{2}+
2\sin^{2}\theta||\Theta _{\chi }||^{2}\overline{
\nabla \strut }\ln r_{\text{in}}
-2\sin^{2}\theta |U|^{-3/4}\overline{\nabla \strut }\ln
r_{\text{in}}+\sin^{2}\theta||\Theta _{\chi
}||^{2}\overline{\nabla
\strut }\ln U,n_{\overline{\chi^{-}}}
\right\rangle_{\chi^{-}} = \\
-\dfrac{\sqrt{r_{0}^{2}-2M^{\prime}r_{0}+a^{2}}}{|U(r_{0})|^{3/4}}
\int\limits_{r=r_{0}}
\left(\sigma^{3/2}\left(||\Theta_{\vartheta} ||^{2}\left(-(1/2)\dfrac{d\ln U}{dr}+ i(3/2)\dfrac{d
\omega}{dr}+2\dfrac{d\ln r_{\text{in}} }{dr }\right)+
2\dfrac{\partial ||\Theta_{\vartheta} ||^{2}}{\partial r}\right)
-2\dfrac{d\ln r_{\text{in}} }{dr }\right)\sin^{2}\theta d\theta
=\\ -\dfrac{1}{|U(r_{0})|^{3/4}}
\int\limits_{r=r_{0}}
\left(\sigma^{3/2}\left(||\Theta_{\vartheta} ||^{2}\left((1/2)r_{\text{in}}\dfrac{d\ln U}{dr_{\text{in}}}
+ i(3/2)\sqrt{r^{2}-2M^{\prime}r+a^{2}}\dfrac{d
\omega}{dr}-2\right)+
2\sqrt{r^{2}-2M^{\prime}r+a^{2}}\dfrac{\partial ||\Theta
||^{2}}{\partial r}\right) +2\right)\sin^{2}\theta d\theta
\end{multline*}
\normalsize
As $r_{0}\rightarrow \infty,$ $||\Theta_{\vartheta}
||^{2}-1=O(r^{-1}).$ The contribution of the remaining terms
vanish as $r_{0}\rightarrow \infty$ by the hypotheses on $U$ and
Eq.~\eqref{dthetasym}.
\end{proof}

We now integrate Eqs.~(\ref{id},\ref{idminus}) on $\Sigma^{\pm}.$
For both cases the boundary integrals at $r=M+c$ (equivalently
$2r_{\text{in}}=2r_{\text{out}}=c$) are evaluated w.r.t. to the
normal vector pointing in the direction of decreasing $r.$

\begin{lemma} \label{lemma11.4}
Suppose $U=U\left(r\right) $ is globally $C^{1}$ and $ U^{-1}$ is
globally bounded and $(d \ln U/d r)=o( r^{-1}) $ as $r
\rightarrow \infty $ and $U$ is a solution of Eq.~\eqref{cqdp} in
$\Sigma^{+}$ and Eq.~\eqref{cqdm} in $\Sigma^{-}$ such that $(d
\ln U/d r)$ is locally bounded in $(M+c,\infty).$ and fails
to be differentiable at most at finite number of points. Then
\scriptsize
\begin{multline} \label{outfinal}
\iint\limits_{\Sigma^{+}_{2},\overline{\chi} }\left(\mathcal{P}\sigma^{-2}\zeta ^{-1}||\Theta _{\chi }||^{2}
+4||\nabla _{\chi }\Theta _{\chi }||^{2}+
||\Theta_{\chi}||^{2}\sigma^{-2}\zeta^{-1}\left(2\mathcal{A}^{-}\overline{g}^{\theta\theta}+i\mathcal{I}_{\text{out}}\right)
\right)\sin^{2}\theta=\\\lim\limits_{r_{0}\downarrow M+c}\dfrac{1}{|U(r_{0})|^{3/4}}
\int\limits_{r=r_{0}}
\left(\sigma^{3/2}\left(||\Theta_{\vartheta} ||^{2}\left((1/2)r_{\text{out}}\dfrac{d\ln U}{dr_{\text{out}}}- i(3/2)
\sqrt{r^{2}-2M^{\prime}r+a^{2}}\dfrac{d
\omega}{dr}-2\right)\right)+2\right)\sin^{2}\theta d\theta
\end{multline}
\normalsize
The corresponding integral on
$(\Sigma^{-}_{2},\overline{\chi^{-}})$ is
\scriptsize
\begin{multline} \label{infinal}
\iint\limits_{\Sigma^{-}_{2},\overline{\chi^{-}}}\left(\mathcal{P}\sigma^{-2}\zeta ^{-1}||\Theta _{\chi }||^{2}
+4||\nabla _{\chi }\Theta _{\chi }||^{2}+
||\Theta_{\chi}||^{2}\sigma^{-2}\zeta^{-1}\left(2\mathcal{A}^{-}\overline{g}^{\theta\theta}
+\mathcal{R}_{\text{in}}+i\mathcal{I}_{\text{in}}\right)
\right)\sin^{2}\theta= \\
-\lim\limits_{r_{0}\downarrow M+c}\dfrac{1}{|U(r_{0})|^{3/4}}
\int\limits_{r=r_{0}}
\left(\sigma^{3/2}\left(||\Theta_{\vartheta} ||^{2}\left((1/2)r_{\text{in}}\dfrac{d\ln U}{dr_{\text{in}}}
+ i(3/2)\sqrt{r^{2}-2M^{\prime}r+a^{2}}\dfrac{d
\omega}{dr}-2\right)\right) +2\right)\sin^{2}\theta d\theta
\end{multline}
\normalsize
The limiting sets for the integrals in the RHS of both equations
are restricted to the horizon only.
\end{lemma}
\begin{proof} Left hand sides result from the right hand sides of
Eqs.~(\ref{id},\ref{idminus}) for the solutions U of
Eqs.~(\ref{cqdp},\ref{cqdm}) after we modify $Q_{\overline{g}}$ to
$\widehat{Q}_{\overline{g}}$ in Eq.~\eqref{widehatQ} and replace
the latter and $L$ by their averages and use
Eqs.~(\ref{extp},\ref{extm}). Right hand sides come from the right
hand sides of Eqs.~(\ref{boundary},\ref{innerboundary}). We note
that in the right hand sides of
Eqs.~(\ref{outfinal},\ref{infinal}) on the horizon
$\sqrt{r^{2}-2M^{\prime}r+a^{2}}\dfrac{\partial
||\Theta_{\vartheta} ||^{2}}{\partial r}$ drops out because of
Eq.~\eqref{dthetahor}. Axis parts do not contribute in the limit
$r_{0}\downarrow M+c.$ One way to see this is from
Eq.~\eqref{varandg}. We see $\sigma^{3/2}||\Theta_{\vartheta}
||^{2}\sin^{2}\theta
d\theta=X_{\text{K}}^{1/8}\zeta^{-1/4}X^{5/8}f^{-3/4}||\Theta_{\widehat{g}}||^{2}\sin^{2}\theta
d\theta =O\left((\sin\theta)^{-1/4}\right)\sin^{2}\theta d\theta$
\, and so differentiating partially relative to the variable $r$
we get $\sigma^{3/2}(\partial ||\Theta_{\vartheta} ||^{2}/\partial
r)\sin^{2}\theta d\theta
=O\left((\sqrt{r^{2}-2M^{\prime}r+a^{2}})^{-1}(\sin\theta)^{-1/4}\right)\sin^{2}\theta
d\theta.$ The factor $(\sqrt{r^{2}-2M^{\prime}r+a^{2}})^{-1}$
comes from coordinate transformation Eqs.~\eqref{rhoandrtheta}
when we change the partial derivative relative to $r$ to partial
derivatives relative to regular coordinates $\rho,z$ and it get
cancelled by the reciprocal factor in front of $(\partial
||\Theta_{\vartheta} ||^{2}/\partial r)$ in the left hand sides of
Eqs.~(\ref{boundary},\ref{innerboundary}). Since on these parts of
the $r=\text{constant}$ surfaces $\sin\theta$ approaches $0$ in
the limit, taking the limit after integration results $0.$
Continuity of $U$ and the ODE ensures the continuity of
dlnU/dr and hence gives the cancellations of the boundary integrals at finite number of
values of $r$ where $U$ fails to be twice differentiable.
\end{proof}
If there are parts on the axis above the topmost and below the
bottommost poles where $r\downarrow M+c$ and
$\sin\theta\rightarrow 0,$ the proof of the main theorem below
becomes much more difficult and we shall need to consider the
expression for $\Omega_{\text{K}}$ in details. We include this
expression which can be checked using Eqs.~(\ref{kn},\ref{drhodz})
and the formula after Eq.~\eqref{grr}.
\begin{equation}\label{omegak}
\Omega_{\text{K}}=\varrho^{2}\left((r-M-c)(r-M+c)+c^{2}\sin^{2}\theta\right)^{-1}
\end{equation}
We recall $(r-M-c)(r-M+c)=r^{2}-2M^{\prime}r+a^{2}.$ Thus as
$r\downarrow M+c$ and $\sin\theta\rightarrow 0,$
\begin{eqnarray*}
  (\partial \ln
\Omega_{\text{K}} /\partial
\theta) &=&(\partial \ln
\varrho^{2} /\partial
\theta)- 2c^{2}(\sin\theta)(\cos\theta)(r^{2}-2M^{\prime}r+a^{2}+c^{2}\sin^{2}\theta)^{-1} \\
  (\partial \ln
\Omega_{\text{K}} /\partial r)
&=&(\partial \ln
\varrho^{2} /\partial
\theta)- 2(r-M)(r^{2}-2M^{\prime}r+a^{2}+c^{2}\sin^{2}\theta)^{-1}
\end{eqnarray*}
We shall need to estimate the following two integrals respectively
on a $\theta=\text{constant}$ curve and on the curve
$r=M+c+\epsilon$ curve. Here $x=r-M, \alpha>1.$
\begin{multline}
\int\limits_{M+c+\epsilon}^{M+c+\alpha\epsilon}  (\partial \ln
\Omega_{\text{K}}/\partial
\theta)(\sqrt{r^{2}-2M^{\prime}r+a^{2}})^{-1}dr \\
= -2c^{2}\int\limits_{c+\epsilon}^{c+2\epsilon}
 \dfrac{(\sin\theta)(\cos\theta)}{(x^{2}-c^{2}\cos^{2}\theta)\sqrt{x^{2}-c^{2}}}dx+O(\epsilon)
   = 2\left.\left(\arctan{\dfrac{x\tan\theta}{\sqrt{x^{2}-c^{2}}}}\right)
   \right|_{c+\epsilon}^{c+\alpha\epsilon} +O(\epsilon) \\
   =2\arctan{\dfrac{\sqrt{c/(2\alpha)}(1-\sqrt{\alpha})(\tan\theta)/\sqrt{\epsilon}+O(\sqrt{\epsilon})}
   {1+(c/(2\sqrt{\alpha})+O(\epsilon))((\tan\theta)/\sqrt{\epsilon})^{2}}}+O(\sqrt{\epsilon})
   = O((\sin\theta)/\sqrt{\epsilon}) \label{consttheta}
  \end{multline}
  \begin{multline}
  \int\limits_{\theta_{1}}^{\theta_{2}} (\partial \ln
\Omega_{\text{K}}/\partial
r)\sqrt{r^{2}-2M^{\prime}r+a^{2}}d\theta \\ =
-2x\int\limits_{\theta_{1}}^{\theta_{2}}
 \dfrac{\sqrt{x^{2}-c^{2}}}{x^{2}-c^{2}+c^{2}\sin^{2}\theta}d\theta+O(\theta_{2}-\theta_{1})
   = -2\left.\arctan{\dfrac{x\tan\theta}{\sqrt{x^{2}-c^{2}}}}
   \right|_{\theta_{1}}^{\theta_{2}}+O(\theta_{2}-\theta_{1})\\
    = -2\left.\arctan{\dfrac{(c+\epsilon)\tan\theta}{\sqrt{2c\epsilon+\epsilon^{2}}}}
   \right|_{\theta_{1}}^{\theta_{2}}+O(\theta_{2}-\theta_{1})
   = O((\sin\theta_{1}-\sin\theta_{2})/\sqrt{\epsilon}) \label{constr}
\end{multline}
We need these integrals only near the limiting set as $r\downarrow
M+c,\sin\theta\downarrow 0$ on the axis segment above the topmost
and below the bottommost pole in case such segments have finite
length. In the picture below we explain the axis segment above the
topmost pole. It does not matter if the $r=\text{constant}$ curves
shown intersect the $\theta=\text{constant}$ more than once. The
point is in case the segment $AB$

\begin{overpic}[scale=1,tics=10]{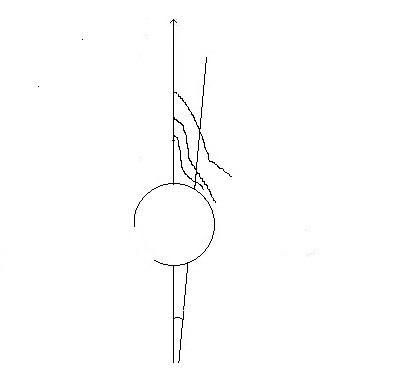}
\put(40,61){$A$}
\put(2,60){$A$ is given by}
\put(2,57){$z=c$ on the
$z$-axis}
\put(60,66){As $\epsilon\downarrow 0$ and $\sin\theta\downarrow 0 $}
\put(60,63){$A^{\prime}$ moves to the point $A$}
\put(60,60){which is $z=c$ on the
$z$-axis}
\put(41,92){$z$-axis}
\put(40,51){$B$}
\put(60,53){$B^{\prime}$ moves to the point $B$}
\put(60,50){which is the topmost pole}
\put(2,50){$B$ is the topmost pole given}
 \put(2,47){by $z=l$ on the $z$-axis}
\put(47,18){$\theta$ angle subtended at the origin $z$-axis}
\put(52,80){$\leftarrow \theta=$constant line}
\put(50,63){$A^{\prime}$}
\put(50,53){$B^{\prime}$}
\put(57,50){$D$}
\put(45,65){$C$}
\put(1,72){$r=M+c+\alpha\epsilon$ curve, $\alpha>1 \rightarrow$}
\put(12,67){$r=M+c+\epsilon$ curve $\rightarrow$}
\put(55,37){Horizon and limiting set $r=M+c.$}
\put(2,37){The limiting set $r=M+c$}
\put(2,34){includes the segment $AB$}
\put(2,31){on the axis but not above $A$}
\end{overpic}

has finite length we can always find a $\theta=\theta(\alpha)$
such that $l_{A}\sin\theta=\text{increment in
}\rho=\int\limits_{c}^{c+\alpha\epsilon}x(x^{2}-c^{2})^{-1/2}dx=\sqrt{2\alpha\epsilon}+O(\epsilon)$
and $l_{B}\sin\theta=\sqrt{2\epsilon}+O(\epsilon)$ where
$l_{A},l_{B}$ are the $z$-cordinates of $A,B.$

\begin{theorem} \label{thm}
There is no analytic multiple black hole solution of
asymptotically flat stationary axisymmetric Einstein-Magnetic
Maxwell equation with non-degenerate event horizon and the single
black hole solution belongs to the Kerr-Newman family.
\end{theorem}
\begin{proof} First we recall that $L_{\text{ave}},Q_{\text{ave}}$ are
independent of $U$ and are fixed functions. They are determined by
the spacetime metric and the known spinor $\Theta_{\widehat{g}}$
which is also determined by the spacetime metric.
$L_{\text{ave}}\rightarrow 1$ as $r\rightarrow
\infty.$
As $r \rightarrow \infty ,$ $Q_{\text{ave}}=O(r^{-1})$ by
Eq.~\eqref{cos2theta}. Thus for large $r,$ $(d\ln U/dr)$ is real
and \scriptsize $\dfrac{d\ln U}{dr
}=\dfrac{-1+\sqrt{1-O(r^{-1})}}{2
\sqrt{r^{2}-2M^{\prime}r+a^{2}}}=O( r ^{-2}).$ \normalsize
So near infinity $U$ is real and $U$ and its derivative have the
appropriate decay needed for \lemref{lemma11.4}. Near $r=M+c,$ $U$
could be imaginary but its real and imaginary parts have the
necessary decay. Near $r=M+c,$ $Q_{\text{ave}}$ is of the form
$C_{1}+C_{2}r $ where $C_{1}$ and $C_{2}$ are real constants. Thus
near $r=M+c,$ $U$ is an approximate solution of
\scriptsize $\dfrac{d\ln U}{dr
}=\dfrac{-C_{3}-C_{4}r}{\sqrt{r^{2}-2M^{\prime}r+a^{2}}}$
\normalsize
for some complex constants $C_{3}$ and $C_{4}.$ This gives
\scriptsize $U\approx
\exp \left( C_{0}+\int
\dfrac{-C_{3}-C_{4}r }{\sqrt{r^{2}-2M^{\prime}r+a^{2}}}dr
\right) =\exp \left( C_{5}-C_{4}\sqrt{r^{2}-2M^{\prime}r+a^{2}}-C_{3}\ln \left(
r-M+\sqrt{r^{2}-2M^{\prime}r+a^{2}}\right) \right).$
\normalsize
$U$ is never zero. Now the question is whether $(d\ln U/dr)$ can
fail to be differentiable at infinite number of values of $r.$ We
already know the behavior of $(d\ln U/dr)$ near $r=M+c$ and for
$r$ sufficiently large.  Away from the
possible zero set of $\Theta$ in the region $r>M+c$ of $3$-manifold, $L$ is analytic. We
consider the points in a compact $r$-interval $[M+c+\epsilon,b]$
where $\epsilon>0$ at which radicals $\sqrt{(2-(3/2)L_{\text{ave}})^{2}-2Q_{\text{ave}}}$ in
Eq.~\eqref{cqdp} (and the corresponding one in Eq.~\eqref{cqdm}) vanish or $L_{\text{ave}}=4/3.$ These
are the likely points where $(d\ln U/dr)$ can fail to be
differentiable although it is regular in an open interval to the
right. If $L_{\text{ave}}(\hat{r})\leq 4/3$ then it follows its
definition Eq.~\eqref{lave} that there is an interval about
$\hat{r}$ such that
$\int\limits_{r=\hat{r}}\sigma^{3/2}||\Theta_{\vartheta}
||^{2}\sin^{2}\theta d\theta>0$ and this integral is analytic in
this interval. So by Eq.~\eqref{lave}, $L_{\text{ave}}$ will be
analytic in this interval. Thus if $L_{\text{ave}}$ attains the
value $4/3$ infinitely many times as $r\rightarrow \hat{r}$ then
$L_{\text{ave}}=4/3$ identically in an interval about $\hat{r}.$
So we assume $L_{\text{ave}}(\hat{r})<4/3$ and ask whether the
radicand can vanish infinitely many times as $r\rightarrow
\hat{r}.$   Similarly for $r>M+c,L_{\text{ave}}(\hat{r})< 4/3,$
in an analytic spacetime $\mathcal{A}(r)$ is an analytic function of $r,$ and so
$Q_{\text{ave}}(r)$ is an analytic function of $r$ when $\mathcal{A}(r)>0.$
So assuming $L_{\text{ave}}(\hat{r})<4/3,$ the radical $\sqrt{(2-(3/2)L_{\text{ave}})^{2}-2Q_{\text{ave}}}$
can vanish only finite number of points.
So $(d\ln U/dr)$ can fail to be
differentiable only for finite number of values of $r$ in the
compact interval $[M+c+\epsilon,b].$ For $L_{\text{ave}}(\hat{r})>4/3,$ $Q_{\text{ave}}(r)$ is not defined at isolated points where
$\int\limits_{r=\hat{r}}\sigma^{3/2}||\Theta_{\vartheta}
||^{2}\sin^{2}\theta d\theta=0$ but these are removable. Thus \lemref{lemma11.4} applies and
we can now add
Eqs.~(\ref{infinal},\ref{outfinal}) to find that the real parts of
the LHS of both equations are zero by virtue of Eq.~\eqref{match1}
provided we take the same value of $|U|$ at $M+c$ for $U$ on both
sides. The real part has nonnegative definite integrand. This
shows $\mathcal{P}=0.$ $\mathcal{P}=0$ gives $\sigma=1$
identically. Thus by asymptotic conditions $X=X_{\text{K}}.$ From
$X=X_{\text{K}}$ alone we cannot conclude that we have a single
black hole. To claim that we have a single black hole we have to
show that the 3-metrics $\widehat{g}$ and $\widehat{g}_{\text{K}}$
are equal. We now prove this fact by showing $\Omega=\Omega_{K}.$
Applying the conformal transformation formula to the metric
$\overline{g}=\Omega\left(d\rho ^{2}+dz^{2}\right)$ we get
$\overline{R}=-\overline{\Delta }\ln
\Omega.$ Similarly and using conformal invariance in $2$-dimension
we have $\overline{R}_{\text{K}}=-\overline{\Delta
}_{\overline{g}}\ln
\Omega_{\text{K}}=-(\Omega/\Omega_{\text{K}})\overline{\Delta }\ln
\Omega_{\text{K}}.$ But Eq.~\eqref{lx} gives
$\overline{R}-(\Omega_{\text{K}}/\Omega)\overline{R}_{\text{K}}=-f_{\text{em}}.$
So we have $\overline{\Delta }\ln
(\Omega_{\text{K}}/\Omega)=-f_{\text{em}}$ which is nonnegative
because in addition to $\mathcal{P}=0$ now we also have
$f_{\text{em}}^{+}=0.$ If we integrate this identity, the boundary
integral at a $r=r_{0}=\text{constant}>M+c$ \ loop is (apart from
the sign) $\int\limits_{r=r_{0}} (\partial \ln (
\Omega_{\text{K}} /\Omega )/\partial
r)\sqrt{r^{2}-2M^{\prime}r+a^{2}}d\theta.$ By asymptotic condition
$\ln(
\Omega_{\text{K}}/\Omega )=O(r^{-1})$ and
$(\partial \ln(
\Omega_{\text{K}}/\Omega )/\partial r)=O(r^{-2})$ as
$r \rightarrow
\infty.$ Thus the boundary integral goes to zero at $\infty.$ Away from the axis the
integral also goes to $0$ as $r\downarrow M+c$ because near the
horizon away from the poles we have  $(\partial \ln
\Omega /\partial r)=O(1)$ and away from the furthest poles and
the adjacent axis segments where $r\downarrow M+c$
 we have $(\partial \ln
\Omega_{\text{K}}/\partial r)=O(1).$ Near the poles $\ln
\Omega=O(\ln\sin\theta)$ and hence $(\partial \ln
\Omega /\partial r)=O(\ln\sin\theta)$ thus the contribution coming from $(\partial \ln
\Omega /\partial r)$ near poles are
$O\left(\int\limits_{\theta_{1}}^{\theta_{2}}\ln\sin\theta d\theta\right)$
which vanishes in the limit as both $\theta_{1},\theta_{2}$ approaches the same value $0$ or $\pi.$
On the axis part given by $r\downarrow
M+c,$
$(\partial \ln
\Omega_{\text{K}} /\partial r )$ is not defined. In fact $\Omega_{\text{K}}^{-1}$ is
zero on these line segments. So we shall integrate the identity
$\overline{\Delta }\ln (\Omega_{\text{K}}/\Omega)=-f_{\text{em}}$
away from these line segments in the following way. In the limit
$r\downarrow M+c,$ the $r=\text{constant}$ curves are coming close
to the inner parts of the axis (and possibly to parts of the axis
adjacent to the topmost or bottommost poles) when there are more
than one black hole. Before we approach each line segment between
two neighboring black holes we take the boundary part to be a
$\theta=\text{constant}$ line segment for $\theta$ close to $0$ or
$\pi.$ On a $\theta=\text{constant}$ line segment the boundary
integral will be (apart from the sign)
$\int\limits_{\theta=\theta_{0}} (\partial \ln (
\Omega_{\text{K}} /\Omega )/\partial
\theta)(\sqrt{r^{2}-2M^{\prime}r+a^{2}})^{-1}dr.$ Now we can arrange that the line $\theta=\text{constant}$ will intersect the $r=\text{constant}$ loop
between two neighboring black holes even number of times. Thus
this integral will be zero between to neighboring intersection
points because limits of the integration will be the same. Between
two neighboring intersection points where the
$\theta=\text{constant}$ line is nearer to the axis we take the
boundary to be the $r=\text{constant}$ loop. Boundary integral
will vanish at this part because now we integrate relative to
$\theta$ between two equal values of $\theta.$ The above argument
does not apply for calculating the contribution of the boundary
integral from the possible parts of the axis adjacent to the
topmost or bottommost poles where, in the limit $r\downarrow M+c,$
a $r=\text{constant}$ curve is coming close to the axis along an
axis segment of nonzero length. Because now we cannot arrange that
a $\theta=\text{constant}$ curve will intersect a
$r=\text{constant}$ curve in even number of times. That is without
the absence of a black in one side this $r=\text{constant}$ curve
cannot cross the $\theta=\text{constant}$ curve to come out. In
the following we only consider the segment over the topmost pole.
In reference to the diagram we show why the length of the segment
$AB$ is zero. Suppose the length of $AB$ is not zero.
$f_{\text{em}}$ is bounded for $r>M+c.$ So on a set of negligible
measure in this region the area integral of $\overline{\Delta
}\ln(\Omega_{\text{K}}/\Omega)$ is negligible. Thus the boundary
line integral along $A^{\prime}B^{\prime}$ differs slightly from
boundary integrals from the curves such as $CB^{\prime}$ and
$A^{\prime}D$ because on the bases $CA^{\prime}$ or $B^{\prime}D$
of the almost triangular regions $CA^{\prime}B^{\prime}$ or
$B^{\prime}DA^{\prime}$ the boundary contributions are becoming
negligible as $\epsilon\rightarrow 0.$ Thus the values of two
integrals in Eqs.~(\ref{consttheta},\ref{constr}) can only differ
slightly as $\epsilon\downarrow 0$ and $\sin\theta\downarrow 0.$
But for the first integral Eq.~\eqref{consttheta} shows that the
limit is not independent of $\alpha$ because
$(\tan\theta)/\sqrt{\epsilon}=\sqrt{2\alpha}/l_{A}+O(\epsilon).$
So $AB$ cannot have positive length and over the topmost pole we
must have $r>M+c.$ Thus the area integral of $f_{\text{em}}$ is
$0.$ Hence $f_{\text{em}}$ is $0.$ Now $\Omega_{\text{K}}^{-1}$ is
a differentiable function vanishing as
$O((r-M-c)(r-M+c)+c^{2}\sin^{2}\theta)$ at the poles and on the
axis where $r\downarrow M+c,$ whereas
$\Omega=O((\sin^{2}\theta)^{-1}$ near the poles. We now integrate
the identity $\overline{\Delta}(\Omega/
\Omega_{\text{K}})
=(\Omega_{\text{K}}/\Omega)|\overline{\nabla \strut
}(\Omega/\Omega_{\text{K}})|^{2}.$ For the boundary integral with
the domain of integration approaching the limiting set
$r\downarrow M+c$ on the axis we follow the previous method of
integration when both $\theta$ and $r$ approach constant value on
the same line segment. We now get $\Omega=\Omega_{K}.$ So we have
a single black hole and arguments of Bunting
\cite{Bunt} or Mazur \cite{Maz} give the uniqueness result.
\end{proof}
\begin{remark} \label{kerrcase}
{\em If $f_{\text{em}}$ vanishes due to the absence of
electromagnetic field we have the Kerr case. Now it is easier to
show $W=W_{\text{K}}$ (before proving $\Omega=\Omega_{\text{K}}$)
by exploiting one of Carter's equations namely $\overline{\nabla
\strut }\left((X^{2}/\rho)\overline{\nabla \strut }(W/X)\right)
=0.$ This equation gives following two identities since we have
$X=X_{\text{K}}.$
\begin{equation*}
  \overline{\nabla \strut }\left( \dfrac{X^{2}}{\rho }\overline{\nabla \strut
}\dfrac{W-W_{\text{K}}}{X}\right) =0,  \qquad \overline{\nabla \strut }
\left( \dfrac{(W-W_{\text{K}})X^{2}}{X\rho }\overline{\nabla \strut
}\dfrac{W}{X}\right) =(X^{2}/\rho)\left|\overline{\nabla \strut
}\dfrac{W-W_{\text{K}}}{X}\right|^{2}
\end{equation*}
Now on the axis
$X^{2}(\partial(W/X)/\partial(\cos\theta))=O(\rho^{2}).$ So the
boundary integrals from the axis for $r>M+c$ in both identitities
vanish. On the even horizon
$X^{2}(\partial(W/X)/\partial(r))=(\sin^{2}\theta)O(1).$ If the
boundary integrals on the horizon as $r\downarrow M+c$ vanish for
the first identity then they will also vanish for the second
identity because $(W-W_{\text{K}})/X$ is constant on each black
hole surface. On the limiting set $r\downarrow M+c$ on the axis
and between the black holes we follow the previous method of
integration when both $\theta$ and $r$ approach constant value on
the same line segment. On the limiting set $r\downarrow M+c$ on
the axis above the topmost or below the bottommost pole we
integrate along a $r=\text{constant}$ curve. For example in the
1st identity the boundary contribution for these two line segments
are proportional to
$\int\limits_{\theta_{1}}^{\theta_{2}}(X^{2}/\rho) (\partial
((W-W_{\text{K}})/X)/\partial
r)\sqrt{r^{2}-2M^{\prime}r+a^{2}}d\theta $ which goes to $0$ as
$\sin\theta_{1},\sin\theta_{2}$ go to $0.$ Similarly in the second
identity too $\rho$ get cancelled and the integral tends to $0.$
Thus all the boundary integrals for the first and hence for both
the identities vanish and we get $W=W_{K}$ from the second
identity. So using $\rho^{2}=VX+W^{2}$ we get $V=V_{K}.$ Now it is
well-known that if $V,W$ are known then $\Omega$ can be solved
uniquely using its asymptotic value. Finally we ask whether in the
Kerr-Newman case one can prove $W=W_{\text{K}}$ before proving
$\Omega=\Omega_{\text{K}}.$ Assuming that $W$ has the same
positivity property as $X$ one expects that by defining
$\sigma=(W/W_{\text{K}})^{1/4}$ one would be able to repeat the
argument for showing $X=X_{\text{K}}.$}
\end{remark}
\section{Conclusion}
We showed that spin-spin interaction cannot hold black holes apart
in stationary equilibrium in an analytic asymptotically flat
axisymmetric spacetime even in the presence of electromagnetic
fields. The way we modified the method step by stem from the
application of positive mass theorem in Bunting and Masood-ul-Alam
\cite{BuntMA} gives us hope that the method can be further modified to drop the axisymmetry assumption.
This would however be a huge program because our method only shows
$X=X_{\text{K}}.$ The equations to show the rest (including the
equation for $\Omega$) are changed without the assumption of
axisymmetry. A more manageable problem the solution of which will
also be a significant progress is as follows. In stead of
Eq.~\eqref{1} one starts with a non-axisymmetric spacetime metric
of the form
$-(V+\epsilon_{tt})dt^{2}+2(W+\epsilon_{t\phi})dtd\phi+2\epsilon_{tx^{A}}dtdx^{A}+
(X+\epsilon_{\phi\phi})d\phi
^{2}+\overline{g}+\epsilon_{\alpha\beta}dx^{\alpha}dx^{\beta}$
where $\epsilon$ is small and has appropriate boundary properties.
In this case one expects that the error from the deviation from
axisymmetry can be absorbed in a modified $Q_{\text{ave}}$ having
the required boundary properties so that our method would work.

\section{Appendix}
We outline the proofs of \lemref{lemma7} and \lemref{newlemma100}.
If necessary further details can be found in Appendix I of
\cite{MA2}. We denote $\nabla_{g_{1}}$by
$\overset{_{1}}{
\nabla}$ and $\nabla_{g_{2}}$ by $\overset{_{2}}{
\nabla }.$ Let $\{ e(1)^{i}\} _{i=1,2,3}$ be
orthonormal frame field of one forms for
$g_{1}=\overline{G}+f_{1}d\phi ^{2}. $ Let $\{ e(2)^{i}\} $ be the
corresponding orthonormal one forms for
$g_{2}=\overline{G}+qf_{1}d\phi ^{2}.$ We take $e(2)^{\phi
}=\sqrt{qf_{1}}d\phi $ because $ qf_{1}\left\langle d\phi ,d\phi
\right\rangle _{g_{2}}=qf_{1}g_{2}^{\phi
\phi }=1.$ Similarly $e(1)^{\phi }=\sqrt{f_{1}}d\phi .$ Thus
$\sqrt{q}e(1)^{\phi }=e(2)^{\phi }.$ Note $e(1)^{A}=e(2)^{A},$
$A=1,2.$ Corresponding orthonormal frame field of vectors are
$e(1)_{A}= e(2)_{A},$ $A=1,2.$ $\dfrac{1}{\sqrt{q}}e(1)_{\phi
}=e(2)_{\phi }.$ $e(2)_{A}=\sqrt{\overline{G}^{11}}\dfrac{\partial
}{\partial x^{A}},$ since $\left\langle e(2)_{A},e(2)
_{B}\right\rangle _{g_{2}}=\delta _{AB}$ but $\left\langle
\dfrac{\partial }{
\partial x^{A}},\dfrac{\partial }{\partial x^{A}}\right\rangle _{g_{2}}=
\overline{G}_{AB}.$ However at a single point we can arrange
$\overline{G}_{AB}$ to be $\delta _{AB}.$ We also note that
$\left\langle e(2)_{\phi },e(2)_{\phi }\right\rangle
_{g_{2}}=qf_{1}\left( e(2)_{\phi }\right) ^{\phi }\left(
e(2)_{\phi }\right) ^{\phi }=f_{1}\left( e(1)_{\phi }\right)
^{\phi }\left( e(1)_{\phi }\right) ^{\phi }=\left\langle
e(1)^{\phi },e(1)^{\phi }\right\rangle _{g_{1}}.$ $\left\langle
\dfrac{\partial }{\partial \phi },\dfrac{\partial }{\partial
\phi }\right\rangle _{g_{2}}=qf_{1}\Rightarrow $ $\overset{_{2}}
{e}_{\phi }=\dfrac{1}{\sqrt{qf_{1}}}\dfrac{\partial }{\partial
\phi }.$ $\left\langle \dfrac{\partial }{\partial \phi
},\dfrac{\partial }{\partial
\phi }\right\rangle _{g_{1}}=f_{1}\Rightarrow e(1)
_{\phi }=\dfrac{1}{\sqrt{f_{1}}}\dfrac{\partial }{\partial \phi
}.$ First we compute the Christoffel symbols. Let
$\Lambda_{AD}^{C}$ be the Christoffel symbols of $\overline{G}.$
Let $\overset{ _{1}}{\Gamma }_{\beta \gamma }^{\mu }$ and
$\overset{_{2}}{\Gamma }_{\beta \gamma }^{\mu }$be the Christoffel
symbols of $g_{1}$ and $g_{2}.$ These symbols are w.r.t. $\left\{
x^{A},\phi \right\} $ coordinates. They are not the connection
coefficients related to the one frame fields.
\begin{equation*}
\overset{_{2}}{\Gamma }_{\phi \phi }^{\phi }=0, \, \overset{_{2}}{\Gamma }_{AB}^{\phi }=0,
\, \overset{_{2}}{\Gamma }_{B\phi }^{A}=0, \,
\overset{_{2}}{\Gamma }_{\phi A}^{\phi }=\dfrac{1}{2}\dfrac{
\partial \ln \left( qf_{1}\right) }{\partial x^{A}}, \, \overset{_{2}}{\Gamma }_{\phi \phi }^{A}=-\dfrac{1}{2}\dfrac{\partial \left( qf_{1}\right) }{\partial x^{A}},
\overset{_{2}}{\Gamma }_{BC}^{A}=\Lambda_{BC}^{A}
\end{equation*}
\begin{equation*}
\overset{_{1}}{\Gamma }_{\phi \phi }^{\phi }=0, \, \overset{_{1}}{\Gamma }_{AB}^{\phi }=0,
\, \overset{_{1}}{\Gamma }_{B\phi }^{A}=0,
\, \overset{_{1}}{\Gamma }_{\phi A}^{\phi }=\dfrac{1}{2}\dfrac{
\partial \ln f_{1}}{\partial x^{A}}, \, \overset{_{1}}{\Gamma }_{\phi \phi }^{A}
=-\dfrac{1}{2}\dfrac{\partial f_{1}}{\partial x^{A}},
\,
\overset{_{1}}{\Gamma }_{BC}^{A}=\Lambda_{BC}^{A}
\end{equation*}
Now we calculate the connection coefficients $\overset{_{2}}{C}$
and $\overset{_{1}}{ C}$ in the frame fields $\{ e(2)_{i}\} $ and
$\{ e(1) _{i}\} $ for the two metrics respectively. On
$\overset{_{2}}{C}$ and $
\overset{_{1}}{C}$ the indices refer to the frame fields not coordinates.
\begin{eqnarray*}
  &\overset{_{1}}{C}_{\phi \phi \phi } = \langle
e(1)_{\phi },
\overset{_{1}}{\nabla }_{e(1)_{\phi }}e(1)_{\phi
}\rangle _{g_{1}}=0, \quad \overset{_{1}}{C}_{\phi \phi
A}=\left\langle e(1) _{\phi },\overset{_{1}}{\nabla }_{e(1)_{\phi
}}e(1)_{A}\right\rangle _{g_{1}}=(1/2) \dfrac{\partial
\ln f_{1}}{\partial x^{A}},\\
  &\overset{_{1}}{C}_{\phi AB}=\left\langle e(1)_{\phi
},\overset{_{1}}{\nabla }_{e(1)_{A}}e(1)_{B}\right\rangle
_{g_{1}}= 0,\quad
\overset{_{1}}{C}_{A\phi B}=\left\langle e(1)_{A},\overset{_{1}
}{\nabla }_{e(1)_{\phi }}e(1)_{B}\right\rangle _{g_{1}}=0,\\
&\overset{_{1}}{C}_{A\phi \phi }=\left\langle e(1)
_{A},\overset{_{1}}{\nabla }_{e(1)_{\phi }}e(1)_{\phi
}\right\rangle _{g_{1}}=-(1/2) \dfrac{\partial
\ln f_{1}}{\partial x^{A}},\quad
\overset{_{1}}{C}_{AB\phi }=\left\langle e(1)_{A},\overset{_{1}
}{\nabla }_{e(1)_{B}}e(1)_{\phi }\right\rangle _{g_{1}}=0,\\
&\overset{_{1}}{C}_{ABC}=\left\langle e(1)_{A},\overset{_{1}}{
\nabla }_{e(1)_{B}}e(1)_{C}\right\rangle
_{g_{1}}= \Lambda_{ABC}+
\overline{G}_{AC}\dfrac{\partial  }{\partial x^{B}}.
  \end{eqnarray*}

We get $\overset{_{2}}{C}$ replacing $f_{1}$ in expressions for
$\overset{_{1}}{C}$ by $qf_{1}.$
\begin{equation*}
\overset{_{2}}{C}_{\phi \phi \phi }=0=\overset{2}{C}_{\phi AB}=\overset{_{2}
}{C}_{AB\phi }, \, \overset{_{2}}{C}_{\phi \phi A}=(1/2) \dfrac{
\partial \ln \left( qf_{1}\right) }{\partial x^{A}}=-
\overset{_{2}}{C}_{A\phi \phi }, \, \overset{_{2}}{C}_{ABC}=\overset{_{1}}{C}_{ABC}
\end{equation*}
In the following Clifford multiplication by $e(1) ^{i}$and $\
e(2)^{i}$ are denoted by $\cdot. $ Distinction is not necessary
because it is multiplication by the same matrix. Clifford relation
is $e(1) ^{i}\cdot e(1) ^{k}+e(1) ^{k}\cdot e(1)
^{i}=-2\delta^{ij}.$ For the $SU(2)$ spinor $\xi\in \c^{2},$ using
$
  \overset{_{1}}{\nabla }_{e(1)_{k}}\xi =e(1)_{k}\left( \xi
\right) -\dfrac{1}{4}\left\langle e(1)
_{i},\overset{_{1}}{\nabla }_{e(1)_{k}}e(1) _{j}\right\rangle
_{g_{1}}e(1)^{i}\cdot e(1)^{j}\cdot \xi,$ we get
\begin{eqnarray*}
 \overset{_{1}}{\nabla }_{e(1)_{B}}\xi &=&e(1) _{B}\left( \xi
\right) -\dfrac{1}{4}\overset{_{1}}{C}_{ijB}e(1)
^{i}\cdot e(1)^{j}\cdot \xi
\\ &=&e(2)_{B}\left( \xi \right)
+(1/4)\overset{_{1}}{C} _{\phi \phi B}\xi
-(1/4)\overset{_{1}}{C}_{ACB}e(1)^{A}
\cdot e(1)^{C}\cdot \xi   \\
&=&e(2)_{B}\left( \xi \right) +\dfrac{1}{4}\overset{_{2}}{C}
_{\phi
\phi B}\xi
-(1/4)\overset{_{2}}{C}_{ACB}e(2)^{A}
\cdot e(2)^{C}\cdot \xi -(1/8)\left( \dfrac{\partial \ln q}{\partial x^{B}}
\right) \xi  \\
&=&\overset{_{2}}{\nabla }_{e(2)_{B}}\xi -(1/8)\left(
\dfrac{\partial \ln q}{\partial x^{B}}\right) \xi
\end{eqnarray*}
Similarly for $e(2)_{\phi }\left( \xi \right) =0 =e(1)_{\phi
}\left(\xi\right),$
\begin{eqnarray*}
\overset{_{1}}{\nabla }_{e(1)_{\phi }}\xi &=&e(1)_{\phi }\left( \xi \right)
-(1/4)\overset{_{1}}{C}_{ij\phi } e(1)^{i}\cdot e(1)^{j}\cdot
\xi
=(1/4)\left(
\dfrac{\partial \ln f_{1}}{\partial x^{A}}\right)
e(1)^{A}\cdot e(1)^{\phi }\cdot
\xi \\ &=&(1/4)\left(
\dfrac{\partial \ln
\left( qf_{1}\right) }{\partial x^{A}}
\right) e(1)^{A}\cdot e(1)^{\phi }
\cdot \xi -(1/4) \left( \dfrac{
\partial \ln q}{\partial x^{A}}\right) e(1)^{A}
\cdot e(1)^{\phi }\cdot \xi  \\ &=&\overset{_{2}}{\nabla }_{ e(2)_{\phi }}\xi-(1/4)
\left( \dfrac{\partial \ln q}{\partial x^{A}}\right) e(2)^{A}
\cdot e(2)^{\phi }\cdot \xi
\end{eqnarray*}
 So $D_{g_{2}}\xi =D{g_{1}}\xi
+\dfrac{3}{8} \left( \dfrac{
\partial \ln q}{\partial x^{B}}\right) e(2)_{B}\cdot \xi $ giving
$D_{g_{2}}\left( q^{-\dfrac{3}{8}}\xi
\right) =q^{-\dfrac{3}{8}}D_{g_{1}}\xi .$ This proves
\lemref{lemma7}.

To prove \lemref{newlemma100} we also need
\begin{equation*}
\overset{_{2}}{\nabla }_{
e(2)_{\phi }}\xi =\dfrac{1}{4}\left( \dfrac{\partial \ln
\left( qf_{1}\right) }{\partial x^{A}}
\right) e(2)^{A}\cdot e(2)^{\phi }
\cdot \xi
\end{equation*}
 Then
\scriptsize
\begin{multline}\label{a1}
  ||\overset{_{1}}{\nabla }\xi||^{2}=  ||\overset{_{2}}{\nabla
}\xi||^{2}+\sum\limits_{B}\left(-(1/8)\left(
\dfrac{\partial \ln q}{\partial
x^{B}}\right)\left(\left\langle\overset{_{2}}{\nabla
}_{e(2)_{B}}\xi,\xi\right\rangle+\left\langle\xi,\overset{_{2}}{\nabla
}_{e(2)_{B}}\xi\right\rangle\right)\right)+\dfrac{1}{64}
\left( \dfrac{\partial \ln q}{\partial x^{B}}\right)\left( \dfrac{\partial \ln q}{\partial x^{B}}\right)||\xi||^{2}\\
-(1/16)
\left( \dfrac{\partial \ln q}{\partial x^{A}}\right)
\left( \dfrac{\partial \ln (qf_{1})}{\partial x^{B}}\right)\left(\left\langle
e(2)^{B}
\cdot e(2)^{\phi
}\cdot\xi,e(2)^{A}
\cdot e(2)^{\phi
}\cdot\xi\right\rangle+\left\langle e(2)^{A}
\cdot e(2)^{\phi }\cdot
\xi,e(2)^{B}
\cdot e(2)^{\phi
}\cdot\xi\right\rangle\right)\\+(1/16)
\left( \dfrac{\partial \ln q}{\partial x^{A}}\right)
\left( \dfrac{\partial \ln q}{\partial x^{B}}\right)\left\langle e(2)^{A}
\cdot e(2)^{\phi }\cdot
\xi,e(2)^{B}
\cdot e(2)^{\phi }\cdot
\xi\right\rangle
\end{multline}
\normalsize
Now $\left\langle e(2)^{B}
\cdot \xi,e(2)^{A}
\cdot \xi\right\rangle+\left\langle e(2)^{A}
\cdot \xi,e(2)^{B}
\cdot \xi\right\rangle=2||\xi||^{2}.$
Also $\left\langle e(2)^{A}
\cdot \xi,e(2)^{B}
\cdot \xi\right\rangle=-\left\langle e(2)^{B}
\cdot e(2)^{A}
\cdot \xi,\xi\right\rangle=2\delta^{AB}||\xi||^{2}-\left\langle e(2)^{B}
\cdot \xi,e(2)^{A}
\cdot \xi\right\rangle.$
 So Eq.~\eqref{a1} gives
\begin{equation}\label{b9}
  ||\overset{_{1}}{\nabla }\xi||^{2}=  ||\overset{_{2}}{\nabla
}\xi||^{2}-\dfrac{1}{8}\left\langle
\overline{\nabla \strut} \ln q,\overline{\nabla \strut}
||\xi||^{2}\right\rangle_{G}-\dfrac{3}{64} |\overline{\nabla
\strut} \ln q|^{2}_{G}||\xi||^{2} -\dfrac{1}{8}\left\langle
\overline{\nabla \strut} \ln q,\overline{\nabla \strut} \ln
f_{1}\right\rangle_{G}||\xi||^{2}
\end{equation}
We take
\begin{eqnarray*}
  g_{1} &=& \chi=\sigma^{2}\zeta\overline{g}+Ufd\phi^{2} \\
  g_{2} &=& \gammaup=\sigma^{2}\zeta\overline{g}+|U|^{-4}Ufd\phi^{2} \\
  f_{1} &=& Uf, \qquad   q =|U|^{-4}
\end{eqnarray*}
Then Eq.~\eqref{b9} gives\\
 $||\nabla_{\chi}\Theta_{\gammaup}||^{2}=
  ||\nabla_{\gammaup}\Theta_{\gammaup}||^{2}+(1/2)\sigma^{-2}\zeta^{-1}\left\langle
\overline{\nabla \strut} \ln |U|,\overline{\nabla \strut}
||\Theta_{\gammaup}||^{2}\right\rangle-(3/4)
\sigma^{-2}\zeta^{-1}\left\langle
\overline{\nabla \strut} \ln |U|,\overline{\nabla \strut}\ln |U|\right\rangle||\Theta_{\gammaup}||^{2}
+(1/2)\sigma^{-2}\zeta^{-1}\left\langle
\overline{\nabla \strut} \ln |U|,\overline{\nabla \strut} \ln
(Uf)\right\rangle||\Theta_{\gammaup}||^{2}$ where norms of vectors
or forms and inner product of vectors and forms are with respect
to 2-metric $\overline{g}.$

We have $\Theta_{\gammaup}=|U|^{1/2}\Theta_{\chi}.$ Recalling
$\overset{_{1}}{\nabla }=\nabla_{\chi}$ we get,\\
$||\nabla_{\chi}\Theta_{\gammaup}||^{2}
=||\nabla_{\chi}\left(|U|^{1/2}\Theta_{\chi}\right)||^{2}
=||(1/2)|U|^{1/2}\left(\nabla
\ln |U|\right)\Theta_{\chi}+|U|^{1/2}\nabla_{\chi}\Theta_{\chi}||^{2}\\
=(1/4)\sigma^{-2}\zeta^{-1}\left\langle
\overline{\nabla \strut} \ln |U|,\overline{\nabla \strut}\ln |U|\right\rangle||\Theta_{\gammaup}||^{2}
+|U|||\nabla_{\chi}\Theta_{\chi}||^{2} +(1/2)|U|\left\langle\nabla
\ln|U|,\nabla ||\Theta_{\chi}||^{2}\right\rangle_{\chi}.$ Thus for some pure imaginary function (or zero) $Im$ we
get
\begin{multline*}
 (1/4)\sigma^{-2}\zeta^{-1}\left\langle
\overline{\nabla \strut} \ln |U|,\overline{\nabla \strut}\ln |U|\right\rangle||\Theta_{\gammaup}||^{2}
+|U|||\nabla_{\chi}\Theta_{\chi}||^{2} +(1/2)|U|\left\langle\nabla
\ln|U|,\nabla ||\Theta_{\chi}||^{2}\right\rangle_{\chi}+Im
\\=||\nabla_{\gammaup}\Theta_{\gammaup}||^{2}+(1/2)\sigma^{-2}\zeta^{-1}\left\langle
\overline{\nabla \strut} \ln |U|,\overline{\nabla \strut}
||\Theta_{\gammaup}||^{2}\right\rangle-(1/4)
\sigma^{-2}\zeta^{-1}\left\langle
\overline{\nabla \strut} \ln |U|,\overline{\nabla \strut}\ln
|U|\right\rangle||\Theta_{\gammaup}||^{2}\\
+(1/2)\sigma^{-2}\zeta^{-1}\left\langle
\overline{\nabla \strut} \ln |U|,\overline{\nabla \strut} \ln
f\right\rangle||\Theta_{\gammaup}||^{2}
\end{multline*}
Now since
$||\Theta_{\chi}||^{2}=|U|^{-1}||\Theta_{\gammaup}||^{2},$ \quad
$\overline{\nabla \strut}
||\Theta_{\chi}||^{2}=-|U|^{-1}||\Theta_{\gammaup}||^{2}\overline{\nabla
\strut}\ln |U|+|U|^{-1}\overline{\nabla \strut}
||\Theta_{\gammaup}||^{2}$ which gives
\begin{equation*}
\left\langle\overline{\nabla \strut} \ln |U|,\overline{\nabla
\strut}||\Theta_{\chi}||^{2}\right\rangle
=-||\Theta_{\chi}||^{2}\left\langle\overline{\nabla \strut}
\ln |U|,\overline{\nabla
\strut}\ln |U|\right\rangle+|U|^{-1}\left\langle
\overline{\nabla \strut}\ln |U|,\overline{\nabla \strut}
||\Theta_{\gammaup}||^{2}\right\rangle
\end{equation*}
Thus we get
\begin{multline*}
 |U|||\nabla_{\chi}\Theta_{\chi}||^{2}+Im
=||\nabla_{\gammaup}\Theta_{\gammaup}||^{2}+(1/2)\sigma^{-2}\zeta^{-1}\left\langle
\overline{\nabla \strut} \ln |U|,\overline{\nabla \strut} \ln
f\right\rangle||\Theta_{\gammaup}||^{2}
\end{multline*}
which gives \lemref{newlemma100}.

\bibliography{}

\end{document}